\documentclass[pre,reprint,floats,amsmath,amssymb,superscriptaddress,floatfixl,aps]{revtex4-2}

\usepackage[utf8]{inputenc}
\usepackage{amsmath,amssymb}

\usepackage{graphicx} %%%
\usepackage{color}
\usepackage{hyperref} %%%
\usepackage[normalem]{ulem}

\usepackage{xr}
\makeatletter

\newcommand*{\addFileDependency}[1]{% argument=file name and extension
\typeout{(#1)}% latexmk will find this if $recorder=0
\@addtofilelist{#1}
\IfFileExists{#1}{}{\typeout{No file #1.}}
}\makeatother

\newcommand*{\myexternaldocument}[1]{%
\externaldocument{#1}%
\addFileDependency{#1.tex}%
\addFileDependency{#1.aux}%
}
\myexternaldocument{supporting}

\usepackage{physics}

\usepackage[dvipsnames]{xcolor}
\usepackage{fourier}

\extrafloats{100}

\begin{document}

\title{Active Diffusion of Self-Propelled Particles in Semi-Flexible Polymer Networks}

% \author{Yeongjin Kim}
% \affiliation[POSTECH]{Department of Physics, Pohang University of Science and Technology (POSTECH), Pohang 37673, Republic of Korea}

% \author{Sungmin Joo}
% \affiliation[POSTECH]{Department of Physics, Pohang University of Science and Technology (POSTECH), Pohang 37673, Republic of Korea}

% \author{Won Kyu Kim}
% \email{wonkyukim@kias.re.kr}
% \affiliation[KIAS]{School of Computational Sciences, Korea Institute for Advanced Study (KIAS), Seoul 02455, Republic of Korea}

% \author{Jae-Hyung Jeon}
% \email{jeonjh@postech.ac.kr}
% \affiliation[POSTECH]{Department of Physics, Pohang University of Science and Technology (POSTECH), Pohang 37673, Republic of Korea}
% \alsoaffiliation[APCTP]{Asia Pacific Center for Theoretical Physics, Pohang 37673, Republic of Korea}

\author{Yeongjin Kim}
%\email{author1@email.com}
\affiliation{Department of Physics, Pohang University of Science and Technology (POSTECH), Pohang 37673, Republic of Korea}

\author{Won Kyu Kim}
\email{wonkyukim@kias.re.kr}
\affiliation{School of Computational Sciences, Korea Institute for Advanced Study (KIAS), Seoul 02455, Republic of Korea}

\author{Jae-Hyung Jeon}
\email{jeonjh@postech.ac.kr}
\affiliation{Department of Physics, Pohang University of Science and Technology (POSTECH), Pohang 37673, Republic of Korea}
\affiliation{Asia Pacific Center for Theoretical Physics (APCTP), Pohang 37673, Republic of Korea}

%\date{\today}

% \begin{figure}
% \includegraphics[height=3.5cm, width=8.3cm]{for_Table_of_Contents_use_only.eps}\par
% {for Table of Contents use only}
% \end{figure}

\begin{abstract}
Mesh-like structures, such as mucus gel or cytoskeleton networks, are ubiquitous in biological systems. These intricate structures are composed of cross-linked, semi-flexible bio-filaments, crucial to numerous biological processes.
In many biological systems, active self-propelled particles like motor proteins or bacteria navigate these intricate polymer networks.
In this study, we develop a computational model of three-dimensional cubic-topological, swollen polymer networks of semi-flexible filaments. We perform Langevin dynamics simulations to investigate the diffusion of active tracer particles navigating through these networks. 
By analyzing various physical observables, 
we investigate the effects of mesh-to-particle size ratio, P{\'e}clet number of active particles, and bending stiffness of the polymer networks upon active trapped-and-hopping diffusion of the tracer.
When the tracer size is equal to or larger than the mesh size, the polymer stiffness substantially enhances trapping while suppressing the hopping process. Notably, the mean trapped time exhibits an exponential growth law to the bending stiffness with an activity-dependent slope. 
An analytic theory based on the mean first-passage time of active particles in a harmonic potential is developed. Our findings deepen the comprehension of the intricate interplay between the polymer's bending stiffness, tracer size, and the activity of tracer particles. This knowledge can shed light on important biological processes, such as motor-driven cargo transport or drug delivery, which hinge on the behavior of active particles within biological gels.
\end{abstract}

%\pacs{}

\maketitle

\section{Introduction}\label{sec1}

Biopolymer networks of semi-flexible filaments having a large bending persistent length are ubiquitous in biological cells, such as collagen, actin, microtubules, and DNA networks~\cite{duarte2021microrheology}.
In many biosystems, the flexural stiffness of biopolymers is a critical physical component that significantly influences various biological properties. 
For instance, the stiffness of the cellular environment determines the morphing of a stem cell~\cite{engler2006matrix}. Moreover, semi-flexibility plays a vital role in biopolymer network topology and cosolute partitioning~\cite{kim2017cosolute}. 
The stiffness of the mucus layer is important in its physiological functions where the mucus has the ability to restore its initial stiffness through biological processes~\cite{wagner2018mucins}. It is also known that the stiffness of the extracellular matrix plays a vital role in tumor cell migration~\cite{pathak2012independent}.

Living systems are highly dynamic and active, as they utilize components that convert chemical energies (e.g., ATPs) from the environment into mechanical ones, driving the system out of equilibrium. The entities converting the energies to motion can be regarded as active or self-propelled particles~\cite{ramaswamy2017active, schweitzer2003brownian, bechinger2016active}. Prominent examples include motor proteins such as kinesin and dynein and microswimmers like E. coli bacteria \cite{genkin2017topological, berg2004coli, bhattacharjee2019confinement, elgeti2015physics}. The nonequilibrium diffusion dynamics of various active particles have been extensively explored both experimentally and theoretically~\cite{ramaswamy2010mechanics, marchetti2013hydrodynamics}. Beyond the single-particle dynamics, quantitative understanding of the active particles in geometric or potential confinement, polymeric environments, and other complex environments is a currently keen interest~\cite{tailleur2009sedimentation, pototsky2012active,maggi2014generalized, qi2020enhanced, theeyancheri2022silico, bhattacharjee2019bacterial, wu2020medical, wang2022systemic, moore2023active, bechinger2016active, abaurrea2020autonomously, olsen2021active, breoni2020active, shaebani2014anomalous, rizkallah2022microscopic, perez2021impact, bhattacharjee2019confinement, brun2019effective, chepizhko2013diffusion, dehkharghani2023self, luo2023simulation, chopra2022geometric, cao2021chain, kumar2023dynamics}.

The thermal diffusion of Brownian tracers in a polymer network has been extensively investigated. These studies reveal that various diffusion dynamics can emerge depending on the physical conditions of mesh-to-particle size ratio, topology and stiffness of the network, and interactions between the polymer and particles~\cite{godec2014collective, zhou2009brownian, kumar2019transport, licinio1997anomalous, lu2022double, lu2021potential, martin2019review, quesada2022coarse, zhao2022molecular, chen2022dissipative, cho2020tracer, quesada2021solute, cai2015hopping, sorichetti2021dynamics, xu2021enhanced}.  In contrast, the dynamics of active tracers in polymer networks remain relatively unexplored. 
Limited attempts have investigated the diffusion of active particles, primarily in polymer solutions or melts~\cite{du2019study, qi2020enhanced, theeyancheri2020translational, theeyancheri2022silico, yuan2019activity, chen2022passive, zottl2023dynamics, sahoo2022transport}, or within random matrices~\cite{moore2023active, olsen2021active, cao2021chain}.
Addressing this research gap, we recently introduced a computational model of active tracer particles in a flexible cubic polymer network. Through explicit simulations, we examined the diffusion of active tracers navigating this flexible polymer environment~\cite{kim2022active}.
It turned out that geometrically trapped mesh-sized tracers can freely diffuse in the polymer network via the activeness-induced hopping mechanism. The trapped times obey an exponential law, with the mean trapped time decreasing with increased activeness (quantified with P\'eclet number). Furthermore, active tracers always reach the regime of Fickian diffusion in the long-time limit in which the diffusivity increases with the P\'eclet number with three distinct scaling regimes.  

Beyond our previous computational study, in this work, we focus on the active diffusion of self-propelled particles in a \textit{semi-flexible} polymer network. 
We explicitly model a semi-flexible polymer network of cubic topology with various bending stiffness, simulating the diffusion of mesh-sized active particles (i.e., active Ornstein-Uhlenbeck particles and active Brownian particles) therein. We find that in semi-flexible networks, the characteristics of active diffusion are critically determined by the geometrical factor including the mesh-to-particle size ratio. We quantify how the bending stiffness impacts the characteristics of trapped and hopping dynamics of active tracers in terms of various physical observables. While the flight length mildly decreases with the bending stiffness, the trapped time exponentially increases with it. We also present an active barrier-crossing theory to quantitatively account for the observed exponential growth law of the trapped time against the bending stiffness. Our computational and analytic studies altogether demonstrate a significant role of the semi-flexibility of the polymer network in the active transport of tracers. 

%This work is structured as follows: In Section~\ref{sec2}, we introduce the simulated semi-flexible polymer network system, elucidate the model for active tracers, and outline the physical observables examined. Section~\ref{sec3} presents our simulation and analytic results. We delve into the properties of active diffusion for three distinct tracer sizes in Sections~\ref{sec3A} through~\ref{sec3C}. In Section~\ref{sec3D}, we introduce a first-passage barrier crossing theory, focusing on an active Ornstein-Uhlenbeck particle confined to a harmonic potential. Section~\ref{sec4} encapsulates and discusses the main results, supplemented by additional simulation results for active Brownian particles. Finally, Section~\ref{sec5} provides conclusions.
The current work is structured as follows. In Sec.~\ref{sec2}, we introduce the simulated semi-flexible polymer network system and also explain the model of active tracers and the physical observables under examination. In Sec.~\ref{sec3}, we present our simulation and analytic results. For three distinct tracer sizes, we investigate the properties of their respective active diffusion in Secs.~\ref{sec3A}--\ref{sec3C}. 
Then, we present a first-passage barrier-crossing theory of an active Ornstein-Uhlenbeck particle confined to a harmonic potential in Sec.~\ref{sec3D}. 
We encapsulate and discuss the main results in Sec.~\ref{sec4}, along with the additional simulation result for active Brownian particles. Finally, conclusions are provided in Sec.~\ref{sec5}.

\section{Model and methods}\label{sec2}

\subsection{The polymer network}

\begin{figure*}
    \includegraphics[width=18cm]{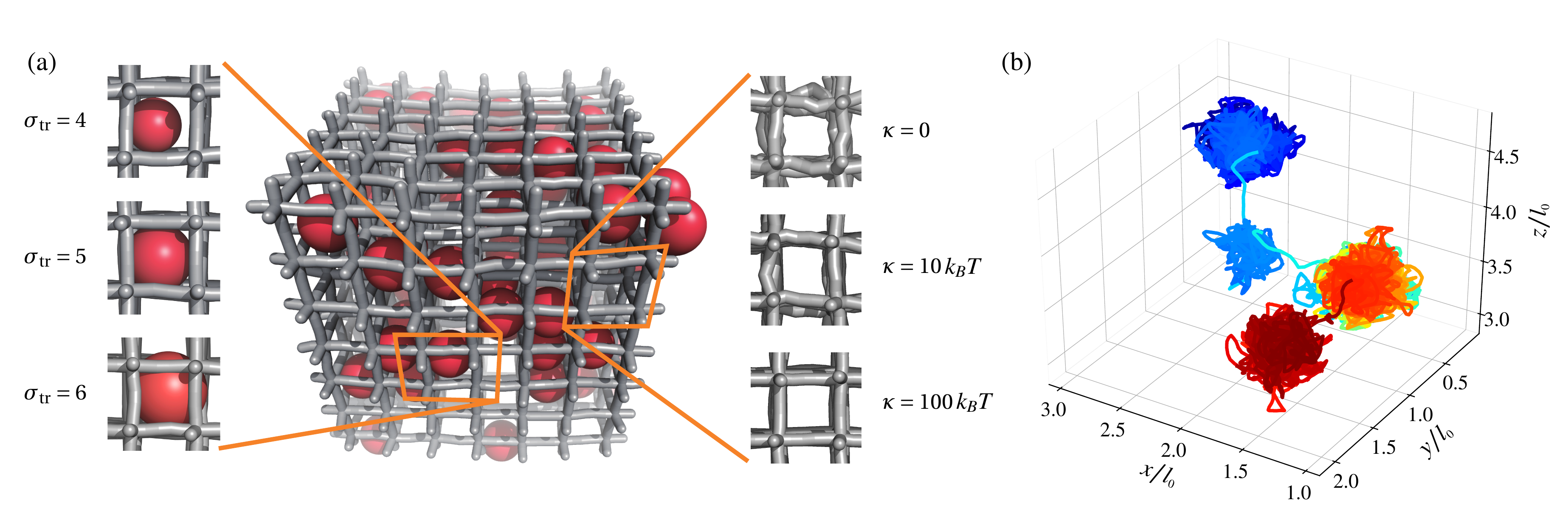}
    \caption{\label{label:Figure01}
    (a) Active tracers (red) in a semi-flexible polymer network gel (gray). The left side illustrates active tracers of varying sizes ($\sigma_{\mathrm{tr}}=4$, $5$, and $6$). On the right side, representative configurations of flexible and semi-flexible polymer networks with different bending stiffnesses ($\kappa=0$, $10~k_B T$, and $100~k_B T$) are shown.
    (b) Representative `trapped-and-hopping' trajectory of a mesh-sized active tracer in the semi-flexible polymer network. The color code depicts the elapsed time (from yellow to blue). We use the tracer diameter $\sigma_{\mathrm{tr}}=5$, the P\'eclet number $\mathrm{Pe}=18$, and the stiffness $\kappa=40~k_B T$.
    }
\end{figure*}

We construct a cubic network of semi-flexible polymers in three dimensions (Fig.~\ref{label:Figure01}a), representing swollen polymer gels that we are interested in. 
The monomers comprising the polymer network are identical with size of $\sigma_0$ and the polymer consists of four monomers, which are interconnected between the nearest cross-linkers. 
The monomers are subject to the bonding potential $U_{\mathrm{bond}}$ and the bending potential $U_{\mathrm{bend}}$. For the bonding potential, we use a bead-spring model, $U_{\mathrm{bond}}(r)=k_b (r-l_b)^2$, where $k_b = 100~{k_B T }\slash\sigma_0 ^2$ is the spring constant~\cite{zhang2019hierarchical}, and $l_b = \sigma_0$ is the bond length. For the bending potential, we adopt a harmonic potential $U_{\mathrm{bend}}(\theta)=\kappa (\theta - \pi)^2$, where $\theta$ is the angle between two bonds, and $\kappa$ is the bending stiffness (modulus) of the polymer network. The stiffness parameter $\kappa$ can be converted into the persistence length ($l_p$) of a semi-flexible polymer via~\cite{kim2017cosolute}
\begin{equation}
    l_p = \frac{2 \kappa \sigma_0 }{k_B T}.
\end{equation}

Figure~\ref{label:Figure01}a illustrates a semi-flexible polymer network ($\kappa=40~k_B T$) and the polymer bending conformations are compared with those with different bending stiffness ($\kappa=0$, $10~k_B T$, and $100~k_B T$). 
It shows that larger stiffness suppresses the fluctuations of the polymer network in the transverse direction, rendering the network more rigid.
In the presence of large fluctuations of the polymer network, the tracer particles therein (red spheres in Fig.~\ref{label:Figure01}a) experience a significant amount of excluded volume, thereby reducing the accessible volume of the tracers within a mesh. Figure~\ref{label:FigureS01} in the Supplementary Material presents the bond-length and angle distributions of the polymer networks for varying $\kappa$.

\subsection{The active tracers and governing equations}

We consider the so-called Active Ornstein–Uhlenbeck (AOU) processes, in which a self-propelled, AOU particle (AOUP) of mass $m$ is fueled by an active force $\mathbf{F}_{\mathrm{A}}$ in the Langevin dynamics, 
\begin{equation}
    \frac{d \mathbf{r}}{dt} = \mathbf{v} , \quad
    m \frac{d\mathbf{v}}{dt} = -\gamma \mathbf{v} + \mathbf{F}_\mathrm{A} + \boldsymbol{\xi}.
    \label{aoup_eom}
\end{equation}
Here, $\mathbf{r}(t)$ is the three-dimensional position, $\mathbf{v}(t)$ is the velocity, $\gamma$ is the friction coefficient, and $\boldsymbol{\xi}=(\xi ^x,\xi ^y, \xi ^z)$ is the thermal white noise with $\langle \xi^\alpha \rangle =0$ and $\left \langle \xi ^\alpha (t) \xi ^\beta (t') \right \rangle = 2 \gamma k_B T \delta _{\alpha \beta} \delta (t-t')$ where $\alpha$, $\beta$ denote the Cartesian component. 

The active force, $\mathbf{F}_\mathrm{A}$, is the OU process: stochastic and governed by the following equation of motion, 
\begin{equation}
    \frac{d \mathbf{F}_\mathrm{A}}{dt} = - \frac{1}{\tau_\mathrm{A}} \mathbf{F}_\mathrm{A}+ \sqrt{\frac{2 \gamma^2 v_p ^2}{3 \tau_\mathrm{A}}} \boldsymbol{\xi}_\mathrm{A}(t),
    \label{aounoise}
\end{equation}
where $\boldsymbol{\xi}_\mathrm{A}= (\xi _\mathrm{A}^x, \xi _\mathrm{A}^y , \xi _\mathrm{A}^z)$ satisfies $\langle \boldsymbol{\xi}_\mathrm{A}\rangle = \mathbf{0}$ and $\left \langle \xi_\mathrm{A}^\alpha (t) \xi_\mathrm{A}^\beta (t') \right \rangle = \delta_{\alpha \beta} \delta(t-t')$, yielding an exponentially decaying correlation in time,
\begin{equation}
    \left \langle  F_{\mathrm{A}} ^\alpha (t) F_{\mathrm{A}} ^\beta (t')\right \rangle = \frac{\gamma ^2 v_p ^2}{3} \delta_{\alpha \beta} e^{-|t-t'|/\tau_\mathrm{A}}.
    \label{eqn_noisecorr}
\end{equation}
Here, $v_p$ represents the strength of propulsion velocity, and $\tau_\mathrm{A}$ represents the directional persistent time. 
This enables us to define the persistence length of AOUP's trajectory as $R_p=v_p \tau_\mathrm{A}$~\cite{nguyen2021active, sevilla2019generalized}.

From Eqs.~\eqref{aoup_eom}--\eqref{eqn_noisecorr}, the mean-square displacement of AOUP is derived as~\cite{nguyen2021active}
\begin{equation}
    \begin{aligned}
& \left \langle \Delta \mathbf{r}^2 (t) \right \rangle_\text{free} =  6 D_{\mathrm{th}}\qty(t - \tau_0 \qty(1-e^{-{t}/{\tau_0}})) \\
%+ \frac{2 v_p^{2} \gamma^{2} \tau^{2} \tau_{\mathrm{A}}}{m^{2}\left(\tau^{2}-\tau_{\mathrm{A}}^{2}\right)} \times \\\\
& + \quad 
2v_p ^2 \tau_{\mathrm{A}} \qty[  \frac{t - \tau_{\mathrm{A}} \qty( 1 - e^{-t/\tau_{\mathrm{A}}})  }{1-\tau_0 ^2 /\tau_{\mathrm{A}} ^2} 
- \frac{\tau_0 ^2}{\tau_{\mathrm{A}} ^2} \frac{t - \tau_0 \qty( 1 - e^{-t/\tau_0})  }{ 1 - \tau_0 ^2/\tau_{\mathrm{A}} ^2 }],
%\left[\left(-1+e^{-\frac{t}{\tau}}\right) \tau^{3}+\left(1-e^{-\frac{t}{\tau_{\mathrm{A}}}}\right) \tau_{\mathrm{A}}^{3}+t\left(\tau^{2}-\tau_\mathrm{A}^{2}\right)\right]
\label{msdfree}
\end{aligned}
\end{equation}
where, $D_{\mathrm{th}}={k_B T}/{\gamma}$ is the long-time diffusivity of a passive Brownian particle, and $\tau_0 = m \slash \gamma$ is the momentum relaxation time.

A quantity of importance representing a degree of activity of AOUPs is the P\'eclet number ($\mathrm{Pe}$), which is defined by
\begin{equation}
    \mathrm{Pe} = \frac{\sigma_{\mathrm{tr}} v_p}{D_{\mathrm{th}}}  = \frac{ 3\pi \eta \sigma_{\mathrm{tr}}^2 v_p }{k_B T}
\end{equation}
where $\sigma_{\mathrm{tr}}$ is the AOUP's size (diameter) and $\eta$ is the viscosity of the medium, which satisfies the Stokes law, $\gamma = 3\pi \eta \sigma_{\mathrm{tr}}$.

\subsection{Non-boned interactions and simulations}
We employ the Lennard--Jones potential for the non-bonded interactions between particles $i$ and $j$,
\begin{equation}
    U_{\mathrm{LJ}}^{i j}\left(r_{i j}\right)=
    \begin{cases}
4 \epsilon_{i j}\left[\left(\frac{\sigma_{i j}}{r_{i j}}\right)^{12}-\left(\frac{\sigma_{i j}}{r_{i j}}\right)^6\right]-U_c & , r_{i j} \leq r_{\mathrm{c}} \\
0 & , r_{i j}>r_{\mathrm{c}}.
\end{cases}
\end{equation}
Here, $r_{ij}$ is the distance between two particles and $\sigma_{ij} = \frac{1}{2} \left ( \sigma_{ii}+\sigma_{jj} \right)$, $r_c=2.5 \sigma_{ij}$ is the cut-off length, and $U_c = 4\epsilon_{ij} \left[ \frac{1}{2.5^{12}} - \frac{1}{2.5^{6}} \right]$. 
We use $\epsilon_{\mathrm{m}~\mathrm{m}} = \epsilon_{\mathrm{m}~\mathrm{c}} = \epsilon_{\mathrm{c}~\mathrm{c}} = \epsilon_{\mathrm{m}~\mathrm{tr}} = \epsilon_{\mathrm{c}~\mathrm{tr}} = 0.1~k_B T$, which represents the repulsive interactions (excluded volumes) between the monomers (m), cross-linkers (c), and tracers (tr)~\cite{kim2019tuning}. For an infinitely dilute condition, we use $\epsilon_{\mathrm{tr}~\mathrm{tr}} = 0 $, similarly to our previous study~\cite{kim2022active}.

We run Langevin dynamics simulations incorporating the system setup and interactions described above. The equation of motion of $i$-th monomer is
\begin{equation}
    \frac{d\mathbf{r}_i}{dt} = \mathbf{v}_i, \quad 
    m\frac{d \mathbf{v} _{i} }{dt} = - \gamma_{i} \mathbf{v} _{i} - \nabla_{i} U _m +\boldsymbol{\xi}_{i}.
\end{equation}
Here, $- \nabla_{i} U_m $ is the force acting on the $i$-th monomer, where the total potential is $U_m = \sum_{ij} \left [ U_{\mathrm{bond}} (r_{ij}) + U_{\mathrm{LJ}} (r_{ij}) \right ] + \sum_{\theta} U_{\mathrm{angle}}(\theta)$, and $\gamma_i = 3\pi \eta \sigma_{i}$ is the friction coefficient.

For the $k$-th AOUPs as tracers, the equation of motion is
\begin{equation}
    \frac{d\mathbf{r}_k }{dt} = \mathbf{v}_k, \quad 
    m \frac{d \mathbf{v}_k}{dt} = - \gamma_k \mathbf{v}_k - \nabla_k U + \mathbf{F}_{\mathrm{A},k} + \boldsymbol{\xi}_k,
\end{equation}
where $U=\sum_{ij} [U_{\mathrm{LJ}} (r_{ij})]$ is the tracer's total LJ potential.

The stochastic active force $\mathbf{F}_{\mathrm{A},k}(t)$ is generated based on Eq.~\eqref{aounoise} for each tracer of index $k$, for which we consider 100 tracers and $\tau_\mathrm{A}=10$.
We use the LAMMPS simulation package~\cite{plimpton1995fast}  to run the above coupled equations of motions with the self-written input scripts.
We use the LJ unit, which sets the unit length $\sigma_0$ and unit time $t_0 = \sqrt{ \left. m \sigma_0 ^2 \right\slash k_B T}$.
The simulation box is of size $35\times 35\times 35$ and periodic, in which seven cross-linkers exist per one linear polymer (total $7\times7\times7$ cross-linkers). We use the time step $\delta t = 0.001$ and the total run time is typically $T = 5\times 10^7~\delta t$. 

\subsection{Mean-square displacement, non-Gaussian parameter, trapped times, and flight lenghts}

From the relative particle position at time $t$, $\mathbf{r}_{\mathrm{rel},i} (t) = \mathbf{r}_i (t) - \mathbf{r}_{\mathrm{cm}} (t)$, where $\mathbf{r}_i (t)$ is the $i$-th particle's position and $\mathbf{r}_{\mathrm{cm}}(t)$ is the center-of-mass position of the system, 
we quantify the characteristic of trapped-and-hopping dynamics by measuring the mean-square displacement (MSD), non-Gaussian parameter (NGP), trapped time ($\tau$), and flight length ($l$).

We define MSD as
\begin{equation}
    \left \langle \Delta \mathbf{r}_{\mathrm{rel}} ^2 (t) \right \rangle = \frac{1}{N} \sum_{i} ^N \frac{1}{T-t}\int_0 ^{T-t} \left(\mathbf{r}_{\mathrm{rel},i}(t' + t) - \mathbf{r}_{\mathrm{rel},i}(t') \right)^2 dt',
\end{equation}
where $N$ denotes the number of tracers ($N=100$). 

From the MSD data, we select a time interval, in which $t > 20$ and the MSD exponent $\alpha = \frac{d \log \langle \mathbf{r}^2_{\mathrm{rel}}(t) \rangle}{d \log t}$ falls within the range $0.95 < \alpha(t) < 1.05$. We then fit the MSD via $\langle \mathbf{r}^2_{\mathrm{rel}}(t) \rangle = {6 D_L t}$, where long-time diffusivity ($D_L$) is the fitting parameter.

We define NGP$(t)$ from the relative position $\mathbf{r}_{\mathrm{rel},i} (t) $, such that NGP in one dimension is 
\begin{equation}
    \mathrm{NGP}(t) = \frac{\langle \Delta x ^4 (t) \rangle}{3\langle \Delta x ^2 (t) \rangle ^2} -1,
\end{equation}
where $x$ is one of the Cartesian components for the relative position, $t$ is the time lag, and the bracket $\langle \cdot \rangle$ represents both time and ensemble average~\cite{kim2022active}.

We employ a bilateral filter to skeletonize obtained trajectories~\cite{tomasi1998bilateral, kim2022active} and collect trapped states and hopping states. Based on this, we calculate average trapped times and flight lengths.

\section{Results}\label{sec3}

%\subsection{trapped-and-hopping mechanism}
From the simulations, we observe that small tracers (either passive or active) of size $\sigma_{\mathrm{tr}} \approx 1$ diffuse inside the polymer network relatively freely. The polymer network acts as a trivial obstacle, thus the tracers' diffusion is suppressed by the percolated geometry, which is qualitatively similar to the free diffusion with decreased mobility~\cite{kim2022active} (Fig.~\ref{label:FigureA01}).

However, we observe different diffusion dynamics when the tracer size becomes comparable to the mesh size of the network ($l_0 \approx 5$), featuring the trapped-and-hopping mechanism~\cite{kim2022active}. 
For instance, Fig.~\ref{label:Figure01}b shows a typical trapped-and-hopping trajectory of a mesh-sized ($\sigma_{\mathrm{tr}}=5$) AOUP tracer (of $\mathrm{Pe}=18$) in a semi-flexible polymer network ($\kappa=40~k_B T$). 

Our main interest lies in the effects of polymer semi-flexibility on such trapped-and-hopping diffusion dynamics of AOUPs.
In the following, we consider three different AOUP tracer sizes: $\sigma_{\mathrm{tr}}=4$, $\sigma_{\mathrm{tr}}=5$, and $\sigma_{\mathrm{tr}}=6$ (Fig.~\ref{label:Figure01}a). We vary $\mathrm{Pe}$ in the range of $0$--$54$ for $\sigma_{\mathrm{tr}}=4$, $0$--$90$ for $\sigma_{\mathrm{tr}}=5$, and $0$--$180$ for $\sigma_{\mathrm{tr}}=6$, respectively.

%=====================================FIG 1=========================================================
%===================================================================================================
\begin{figure*}
\includegraphics[width=18cm]{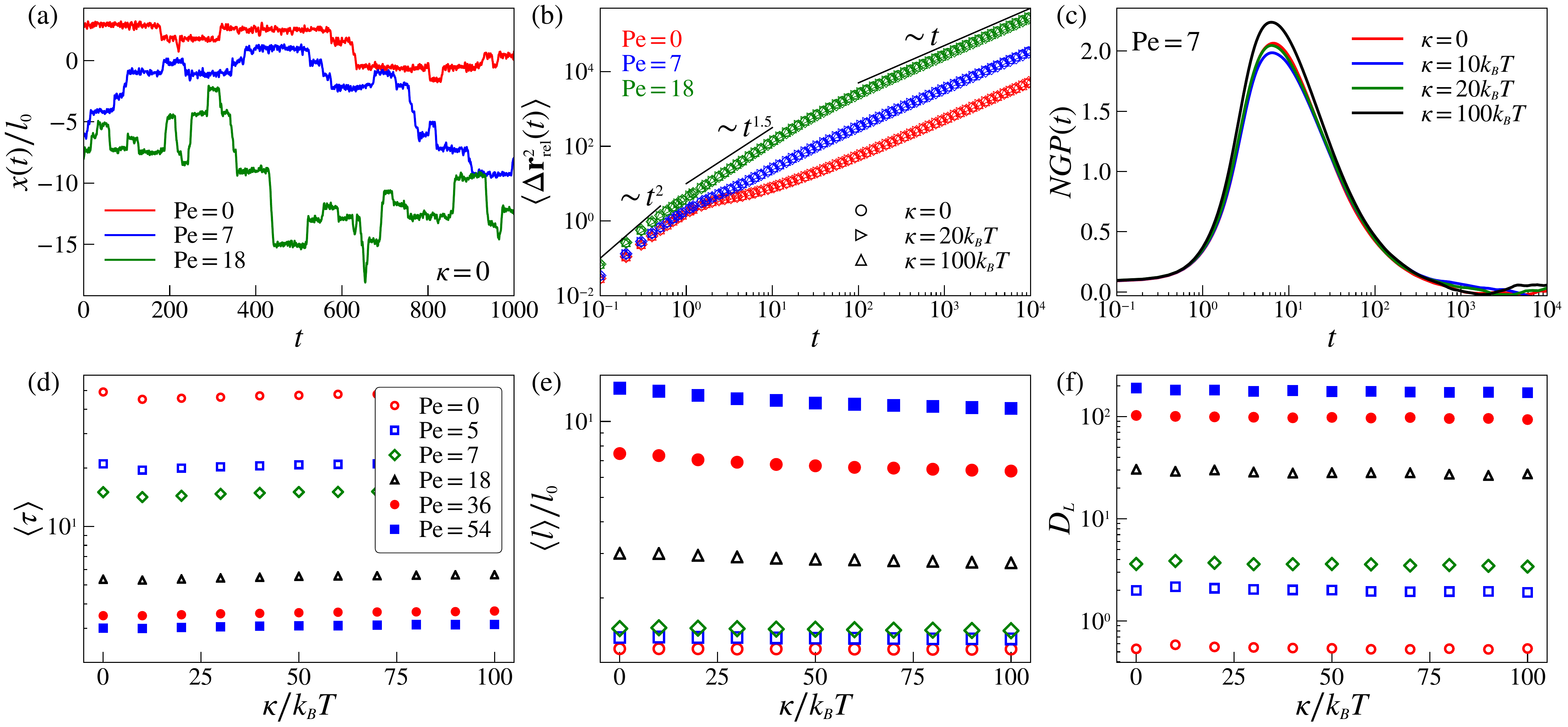}
\caption{\label{label:Figure02} 
(a) Time-dependent position $x(t)$ of an active tracer of size $\sigma_{\mathrm{tr}}=4$ in a flexible ($\kappa=0$) polymer network, for different $\mathrm{Pe}=0$, $7$, and $18$.  
    (b) Mean-square displacement (MSD) of active tracers in polymer networks for different $\mathrm{Pe}$ and $\kappa$. 
    (c) non-Gaussian parameter (NGP) of active tracers with $\mathrm{Pe}=7$ in polymer networks for different $\kappa$. 
    (d) Mean trapped time $\langle\tau\rangle$, (e) mean flight length $\langle l \rangle$, and (f) long-time diffusivity $D_L$ of active tracers as a function of $\kappa$ for different $\mathrm{Pe}$. 
    % (a) The MSD for passive ($\mathrm{Pe}=0$) tracers with size $\sigma_{\mathrm{tr}}=4$. 
    % (b) The MSD for intermediate active ($\mathrm{Pe}=7$) tracers with size $\sigma_{\mathrm{tr}}=4$. 
    % (c) The MSD for strong active ($\mathrm{Pe}=18$) tracers with size $\sigma_{\mathrm{tr}}=4$.
    % (d) The NGP for passive ($\mathrm{Pe}=0$) tracers with size $\sigma_{\mathrm{tr}}=4$. 
    % (e) The NGP for intermediate active ($\mathrm{Pe}=7$) tracers with size $\sigma_{\mathrm{tr}}=4$. 
    % (f) The NGP for strong active ($\mathrm{Pe}=18$) tracers with size $\sigma_{\mathrm{tr}}=4$.
    }
\end{figure*}
%===================================================================================================
%===================================================================================================

%THESE SENTENCES WILL BE GO TO THE APPENDIX
%These features can be also found in the bond-length and -angle distributions as in Fig.~\ref{label:FigureS01}.

% \subsection{$\sigma_{\mathrm{tr}}=1$}

% As in Fig.~\ref{small_msdngp}, the mean square displacement of the small tracers does not show the polymer network stiffness dependency. Also, the non-Gaussian parameters (NGPs) show very small changes. So, in this study, we focus on the tracers whose size is comparable to the mesh size ($\sigma_{\mathrm{tr}}\siml_0 =5\sigma_0$), i.e. $\sigma_{\mathrm{tr}}=4$, $\sigma_{\mathrm{tr}}=5$ and $\sigma_{\mathrm{tr}}=6$.

\subsection{AOUPs with $\sigma_{\mathrm{tr}}=4$}\label{sec3A}

% When the active particle size is comparable to the polymer network mesh size, the transport dynamics are governed by the trapped-and-hopping dynamics~\cite{kim2022active}. (See Fig.~\ref{label:Figure01}b) In this study, we focus on the trapped-and-hopping dynamics of active tracers within semi-flexible polymer networks. 

First, we examine the case in which the tracer size is marginally smaller than the mesh size of the polymer network, i.e., $\sigma_{\mathrm{tr}}=4 \lesssim l_0 = 5$. 
Figure~\ref{label:Figure02}a shows $x(t)$ of the tracer in a flexible polymer network ($\kappa=0$). 
Depending on $\mathrm{Pe}$, the trajectory shows the occasional hopping dynamics, the distance of which increases with larger $\mathrm{Pe}$. Such tendencies are still observed even for semi-flexible networks of $\kappa>0$, i.e., the hopping dynamics is nearly independent of $\kappa$ (data not shown). 

Figure~\ref{label:Figure02}b shows the tracers' MSD$(t)$ for different $\mathrm{Pe}$ and $\kappa$. 
We find that the diffusion dynamics is almost independent of $\kappa$ (different symbols with the same color), implying that the AOUP of $\sigma_{\mathrm{tr}}=4$ is still small and its hopping is subject to minimal contacts with the polymers.
The red symbols depict the MSD of passive tracers ($\mathrm{Pe}=0$), showing the ballistic short-time regime ($\mathrm{MSD}\sim t^2$) before reaching the momentum relaxation time ($t  \lesssim \tau_0 = 1.2 $), and the long-time hopping dynamics ($t  \gtrsim \langle \tau \rangle  \approx 50  $) that is Fickian (MSD$\sim t^1 $). Here, $\langle \tau \rangle$ is the mean trapped time. 
In the intermediate time ($ \tau_0 \lesssim t \lesssim \langle \tau \rangle $), the exponent $\alpha$ of $\text{MSD}(t) \sim t^\alpha$ becomes smaller than unity, signifying the caging effect before reaching the normal diffusion ($\alpha=1$).
% The inset is the zoomed-in MSD at the earlier time with various stiffness, $\kappa$. It shows that the stiffness increases the MSD for the passive tracers in the earlier time \cmtYJ{(See also Supp ???)}. However, in the larger time, there is optimal $\kappa$ that maximizes the MSD \cmtYJ{(See supp?? )}.
The blue symbols represent the MSD of active tracers with $\mathrm{Pe}=7$, showing the almost immediate crossover from the ballistic to Fickian dynamics without any significant intermediate subdiffusive time domain.
This reflects that in the trapped state during the time in the range of $\tau_0 (\approx 1) \lesssim t \lesssim \langle \tau \rangle (\approx 15$), the active force (increasing $\alpha$) and the trapping force (decreasing $\alpha$) become comparable and compete with each other, resulting in the exponent $\alpha$ close to unity. 
When the activeness increases ($\mathrm{Pe}=18$), as shown by the green symbols, the activated hopping dynamics dominates, yielding the trapped time very short ($\langle \tau \rangle \approx 5$). The resultant MSD reveals three temporal regimes divided by the momentum relaxation time ($\tau_0 \approx 1$) and the directional persistent time ($\tau_\mathrm{A}=10$). In the short-time regime ($t\lesssim \tau_0$), similarly to the previous cases, the tracers undergo the ballistic dynamics featuring $\alpha=2$. In the intermediate time ($\tau_0 \lesssim t \lesssim \tau_\mathrm{A}$), however, the competition between the activeness and trapping force yields $\alpha \approx 1.5$. In the long-time regime, the MSD exhibits the Fickian dynamics with $\alpha=1$ eventually.

Figure~\ref{label:Figure02}c depicts the NGP of the tracers with $\mathrm{Pe}=7$. The NGP is maximized at the timescale that the tracer starts to encounter the polymer networks as obstacles. 
Similarly to the MSD, we find that the overall behavior of NGPs barely depends on the polymer network's stiffness. However, the magnitude of the peaks slightly varies depending on $\kappa$, increasing with $\kappa$ for the semi-flexible polymers. This effect is more dramatic for larger tracers, which we discuss further in the following sections.

\begin{figure*}
\includegraphics[width=0.99\textwidth]{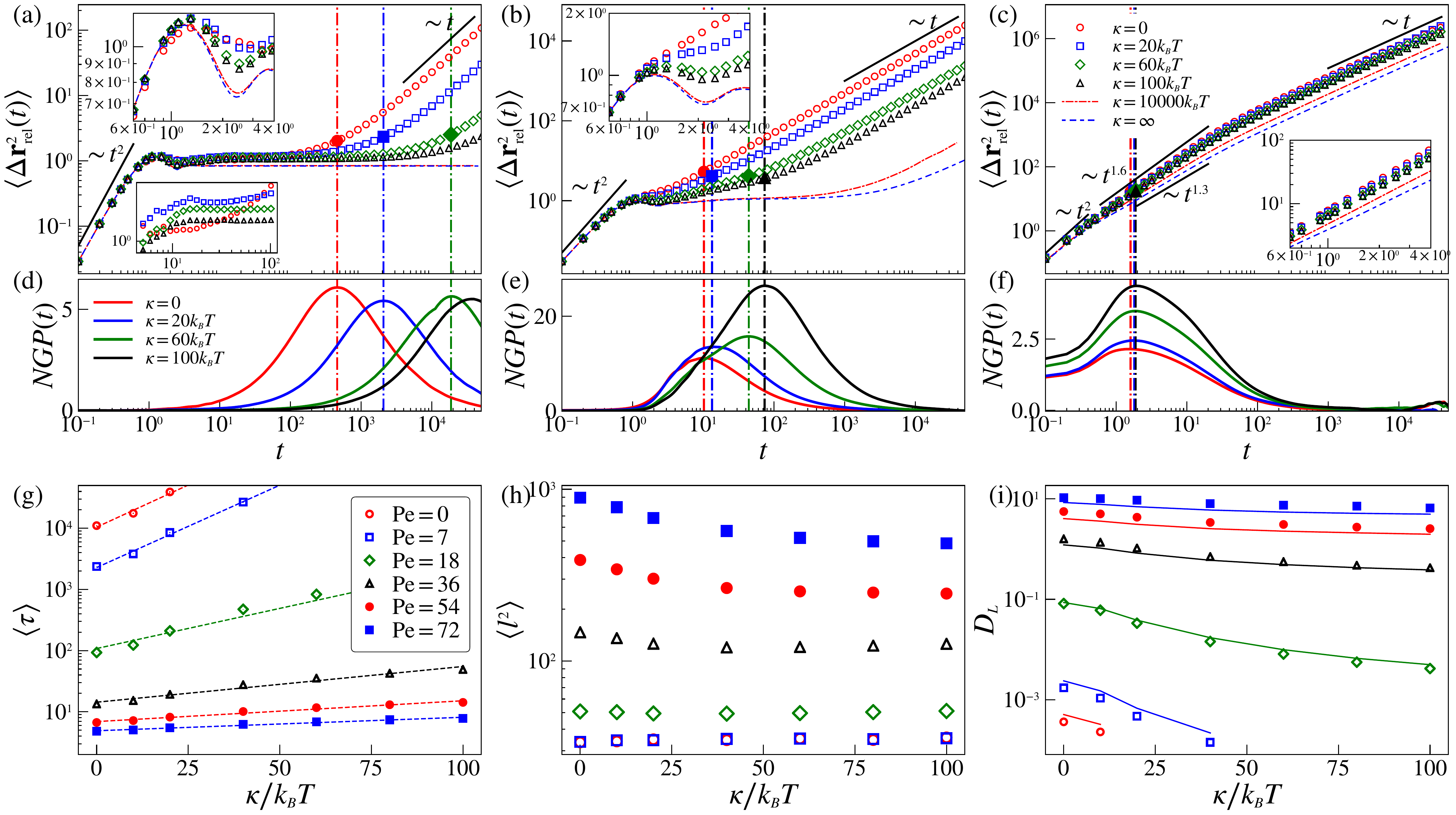}
\caption{\label{label:Figure03} 
MSD of tracers of size $\sigma_{\mathrm{tr}}=5$ for (a) $\mathrm{Pe}=0$, (b) $\mathrm{Pe}=18$, and (c) $\mathrm{Pe}=72$.
NGP for (d) $\mathrm{Pe}=0$, (e) $\mathrm{Pe}=18$, and (f) $\mathrm{Pe}=72$.
(g) Mean trapped time $\langle \tau \rangle$, (h) mean-square flight length $\langle l^2 \rangle$, and (i) long-time diffusivity $D_L$, as a function of $\kappa$, respectively. Solid lines in (i) depict $D_L$ in Eq.~(12).
}
\end{figure*}

Figures~\ref{label:Figure02}d, \ref{label:Figure02}e, and \ref{label:Figure02}f show the mean trapped time, the mean flight length, and the long-time diffusivity, as a function of $\kappa$, respectively. (The distributions of the trapped times and flight lengths are present in Figs.~\ref{label:FigureS02} and \ref{label:FigureS03}).
Those key quantities of our interest are not significantly dependent on the polymer stiffness, in line with the tendency we find from $x(t)$ and MSD.

\subsection{AOUPs with $\sigma_{\mathrm{tr}}=5$}\label{sec3B}

Now we consider AOUPs whose size is equal to polymer network's mesh size ($\sigma_{\mathrm{tr}}=l_0=5$). For a limiting reference, we also consider a network comprising rigid rod-like polymers ($\kappa \rightarrow \infty$), the conformational fluctuation of which vanishes.  
%In Fig.~\ref{label:Figure03}, we present the typical MSDs, NGPs, trapped time distributions, flight length distributions, and the long-time diffusivity of the mesh-sized tracer ($\sigma_{\mathrm{tr}}=5$). 
%As in  the $\sigma_{\mathrm{tr}}=4$ case, all quantities are calculated based on the relative trajectories ($\mathbf{r}_{\mathrm{rel}}$) of the tracer, where the center of mass position of the system is substituted ($\mathbf{r}_{\mathrm{rel},i}(t) = \mathbf{r}_i(t) - \mathbf{r}_{\mathrm{cm}}(t)$). 

%The MSDs of AOUPs with different $\mathrm{Pe}$ are shown in Figs.~\ref{label:Figure03}a--c. 
Figure~\ref{label:Figure03}a shows MSDs of passive tracers ($\mathrm{Pe}=0$) for different $\kappa$. As similarly found in the previous section, we find that the MSD in the short-time ballistic regime ($t\lesssim \tau_0$) varies less sensitively with $\kappa$.
The subsequent time regime exhibits the pronounced caged dynamics in $\tau_0 \lesssim t \lesssim \langle \tau \rangle $, after which the long-time Fickian dynamics reflects the hopping dynamics between network meshes. 

The upper inset in Fig.~\ref{label:Figure03}a highlights two intriguing features of the short-time dynamics. 
One is the enhancement of MSD with increasing $\kappa$ at very short time ($t \lesssim 1$). 
The conformational fluctuation of the flexible polymer network is larger than that of the semi-flexible polymer networks. This implies that the effective free volume accessible to the tracers is larger for stiffer networks. 
Therefore, the polymer networks with larger $\kappa$ enhance the magnitudes of short-time MSDs.
The other notable feature is the bounce-back dynamics, i.e., the oscillatory behavior of MSDs.
This occurs when AOUPs bounce back and forth within a network mesh. We find that the bounce-back dynamics originates from the underdamped dynamics under confinement, consistent with literature~\cite{caprini2021inertial, debnath2021escape}. For the same polymer network system, the overdamped AOUP does not exhibit the bounce-back dynamics (See our supplementary overdamped Langevin dynamics simulation in Fig.~\ref{label:FigureS04} ).
For large $\kappa$, the bounce-back dynamics becomes more pronounced, yielding large oscillatory MSDs.

This tendency changes in the subsequent confined dynamics (see the lower inset in Fig.~\ref{label:Figure03}a). For semi-flexible networks of stiffness up to $\kappa=20~k_B T$, the accessible free volume increases and the MSDs become larger. But for stiffer networks ($\kappa > 20~k_B T$), the shape of the potential of mean force, whose energy barrier is larger than the one of a flexible network ($\kappa=0$), comes into a major play, resulting in decreasing MSDs with $\kappa$ (see Fig.~\ref{label:FigureA02}a for the potential of mean force).

When the activeness of AOUPs is larger ($\mathrm{Pe}=18$), as shown in the inset of Fig.~\ref{label:Figure03}b, we find that the bounce-back dynamics disappear for $\kappa < \kappa^\ast \approx 20~k_B T$, implying that the critical stiffness $\kappa^* (\mathrm{Pe})$ exists, above which the bounce-back dynamics occurs.
In addition, as shown in Fig.~\ref{label:Figure03}b, the short-time dynamics ($t < 1$) barely depends on $\kappa$ and the activated tracers undergo the later escape dynamics easily, resulting in the shortened intermediate confined dynamics. 

When $\mathrm{Pe}$ is very large ($\mathrm{Pe}=72$), as shown in Fig.~\ref{label:Figure03}c, the active force dominates over the trapping force and we observe the similar tendency of MSD $(\alpha > 1)$ found in Fig.~2b. There occurs a crossover between the superdiffusive, intermediate, and Fickian dynamics, divided by the momentum relaxation time $\tau_0$ and the directional persistent time $\tau_\mathrm{A}$, in which the exponent value varies ($\alpha=1.3$--$1.6$) in the intermediate time.  Nevertheless, unlike in the small AOUPs' case, the MSD largely depends on $\kappa$ and its magnitude decreases with $\kappa$.

Below the MSDs, in Figs.~\ref{label:Figure03}d--f, we show NGPs in the same time axes. 
We find that the NGPs have a peak (maximally non-Gaussian) at a certain intermediate time.
The vertical lines depict this peak time of NGP, respectively, for different $\kappa$, and the solid symbols depict the corresponding MSDs at the peak time.
What we first find is that the peak time of NGP increases as $\kappa$ increases.
However, the peak time becomes shorter and less dependent on $\kappa$ for large $\mathrm{Pe}$ (see Figs.~\ref{label:Figure03}e and \ref{label:Figure03}f).
Although the peak time of NGP tends to be almost independent of $\kappa$ for large $\mathrm{Pe}$, the magnitude of peak increases with $\kappa$, unlike NGP for the case of $\mathrm{Pe}=0$.
This means that the non-Gaussian dynamics is dictated by the high competition between the activeness and the trapping force enhanced by increasing $\kappa$, the mechanism of which is well captured by Fig.~\ref{label:Figure03}e at $\mathrm{Pe}=18$ where NGP's magnitude is around five times larger than the other cases (trapping dominant at $\mathrm{Pe}=0$ and activeness dominant at $\mathrm{Pe}=72$).
This feature is also captured by the van-Hove self-correlation functions~\cite{kim2022active} (Fig.~\ref{label:FigureS05}).

\begin{figure*}
    \includegraphics[width=0.95\textwidth]{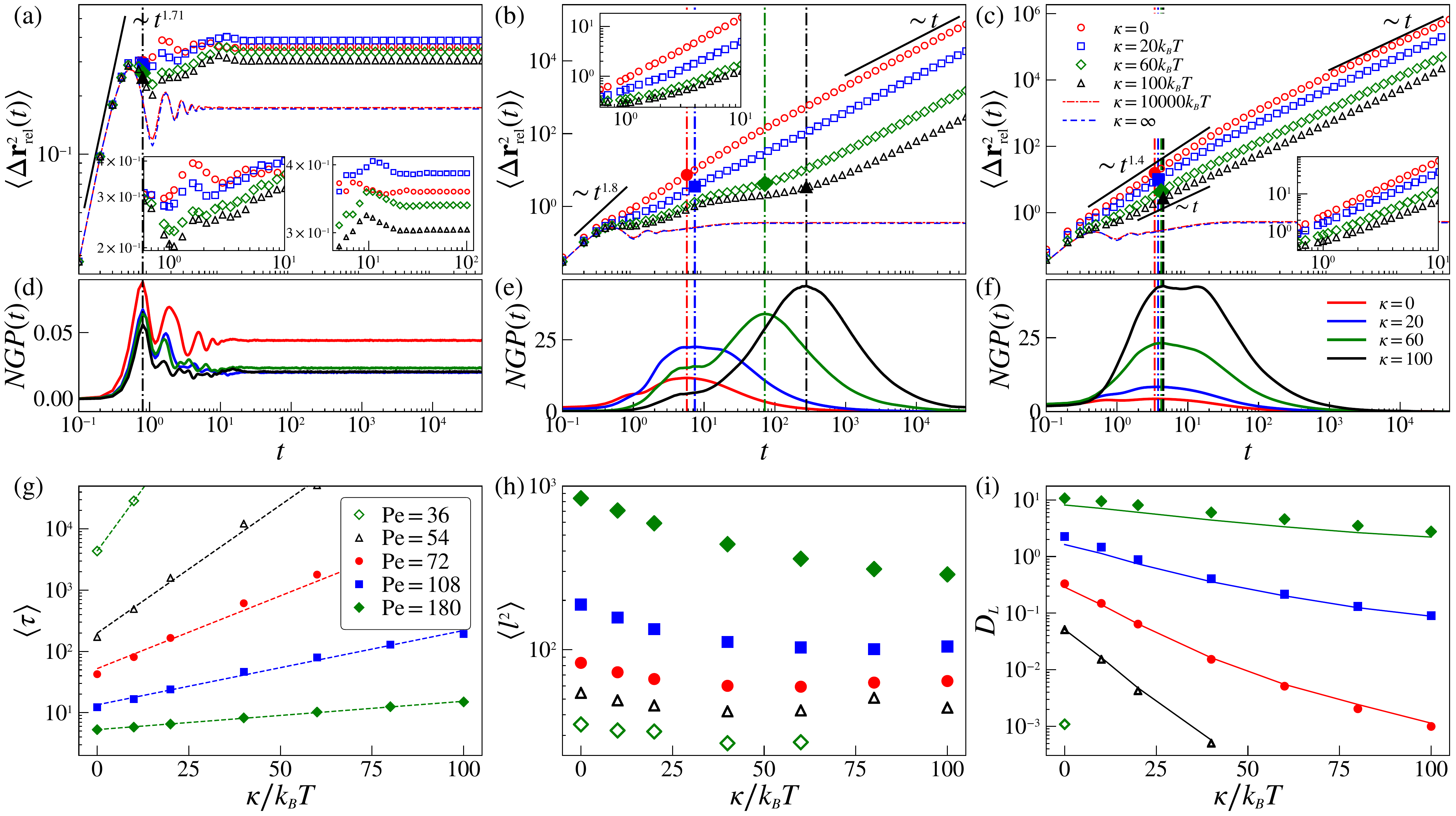}
    \caption{\label{label:Figure04} 
    MSD of tracers of size $\sigma_{\mathrm{tr}}=6$ for (a) $\mathrm{Pe}=0$, (b) $\mathrm{Pe}=72$, and (c) $\mathrm{Pe}=108$.
    NGP for (d) $\mathrm{Pe}=0$, (e) $\mathrm{Pe}=72$, and (f) $\mathrm{Pe}=108$.
    (g) Mean trapped time $\langle \tau \rangle$, (h) mean-square flight length $\langle l ^2 \rangle$, and (i) long-time diffusivity $D_L$, as a function of $\kappa$, respectively.
    Solid lines in (i) depict $D_L$ in Eq.~(12).
    }
\end{figure*}

The trapped-and-hopping dynamics is also characterized by the mean trapped time $\langle \tau \rangle$ and the mean-square flight length $\langle l^2 \rangle$.
Figure~\ref{label:Figure03}g shows $\langle \tau \rangle$ as a function of $\kappa$. In stark contrast to the small tracers ($\sigma_{\mathrm{tr}}=4)$, the mean trapped time of larger tracers depends on $\kappa$ significantly. 
Notably, for the range of $\mathrm{Pe}$ explored, $\langle \tau \rangle$ increases exponentially with the bending stiffness $\kappa$. 
In the plot, the dashed line depicts the best exponential fit to the simulation data. 
The stiffness-induced increase in $\langle \tau \rangle$ is in line with the $\kappa$-dependent tendency in MSD and NGP (Figs.~\ref{label:Figure03}a and \ref{label:Figure03}d).
The slope of the exponential growth tends to decrease as the activeness of the tracer is larger. In the limit of very large $\mathrm{Pe}$, AOUPs can jump to other meshes very easily even in a stiff polymer network in which the mean trapped time does not depend on $\kappa$.
Apart from the mean trapped time, we also examine the distribution of trapped times at $\mathrm{Pe}=0,~12$, and 72 (Fig.~\ref{label:FigureS06}). As in the case of a flexible polymer network~\cite{kim2022active}, the distributions are always exponentially decaying ones regardless of the bending stiffness. Here, the effect of the bending stiffness enters in the characteristic time of the distribution, i.e., $\langle \tau \rangle$.

In Fig.~\ref{label:Figure03}h, we show $\langle l^2 \rangle$ as a function of $\kappa$. 
As expected, the flight length becomes larger with increasing $\mathrm{Pe}$ for all $\kappa$ values investigated. By contrast, the $\kappa$-dependence in $\langle l^2 \rangle$ is nontrivial. 
For small $\mathrm{Pe}$ of $\lesssim 18$, the flight length is almost independent of $\kappa$. In this regime, the mesh-induced trapping force dominates over the active force, so hopping events are restricted to the nearest-neighbor jump only. For larger $\mathrm{Pe}$ of $\gtrsim54$, the active tracer can escape from trapping and jump over multiple mesh distances. This is indeed shown in the flight length distribution plotted in Fig.~\ref{label:FigureS07}. In this case, the hopping active tracer tends to feel increasing polymer friction as the network is stiffer, so $\langle l^2 \rangle$ decreases with $\kappa$. Notably, beyond a certain bending stiffness, the tracer effectively sees a very similar polymer environment against increasing $\kappa$, thus exhibiting a plateau in $\langle l^2 \rangle$. 

In Fig.~\ref{label:Figure03}i, we show the long-time diffusivity $D_L$ as a function of $\kappa$ for different $\mathrm{Pe}$. 
For small $\mathrm{Pe}$, $D_L$ is a decreasing function of $\kappa$, while for large $\mathrm{Pe}$ it becomes nearly independent of $\kappa$ due to the excessively high activeness.
We can quantitatively explain the behavior of $D_L$ using the following relation~\cite{kim2022active}
\begin{equation}\label{eq:DL}
D_L = \frac{1}{6}\frac{\langle l^2 \rangle}{\langle \tau \rangle + \langle \tau_\text{fl} \rangle},  \end{equation}
depicted by the solid lines in Fig.~\ref{label:Figure03}i. 
As $\kappa$ increases, the flight length tends to decrease (Fig.~\ref{label:Figure03}h) while the trapped time increases (Fig.~\ref{label:Figure03}g). In addition, the mean flight time $\langle \tau_\text{fl} \rangle$ is found to be a decreasing function of $\kappa$ (Fig.~\ref{label:FigureA03}a).
Interestingly, at large $\mathrm{Pe}$ values, the magnitude of $\langle \tau_\text{fl} \rangle$ becomes comparable with that of $\langle \tau \rangle$ and their variation tendency is opposite against increasing $\kappa$. Consequently, $\langle \tau \rangle$ and $\langle \tau_\text{fl} \rangle$ offset each other and the denominator in Eq.~\eqref{eq:DL} becomes nearly independent of $\kappa$, resulting in a decreasing $D_L$ with $\kappa$ mainly due to $\langle l^2 \rangle$. 

\subsection{AOUPs with $\sigma_{\mathrm{tr}}=6$}\label{sec3C}

Lastly, we study the case where the AOUP tracer size is slightly larger than the network mesh size ($\sigma_{\mathrm{tr}}=6 \gtrsim l_0 = 5$), as shown in Fig.~\ref{label:Figure01}a. 

Figure~\ref{label:Figure04}a depicts the MSD of passive tracers ($\mathrm{Pe}=0$) for different $\kappa$. In contrast to the previous case ($\sigma_{\mathrm{tr}}=5$), the short-time MSD (in $t < 1$) is found to be independent of the bending stiffness.
This is because the accessible volume for the large tracer is very small regardless of the magnitude of bending stiffness $\kappa$. 
However, at longer times of $1 < t < 10$, the tracers are in contact with the polymers, exhibiting MSD decreasing as $\kappa$ increases (Fig.~\ref{label:Figure04}a, left inset), similarly found in the previous case of $\sigma_\text{tr}=5$.
For $t > 10$, MSD increases with $\kappa$ up to $\kappa=20~k_B T$, beyond which it decreases (Fig.~\ref{label:Figure04}a, right inset).

%\textAdded{The MSDs at $t > 10$ stands for the potential of mean force, shown in Fig.~\ref{label:FigureA02}b. For $\kappa < 20~k_B T$, the width of potential of mean force does not change much, which is reflected as small increases in the right inset. However, for $\kappa > 20~k_B T$, the width of $U(x)$ decreases and the potential becomes stiffer, which is reflected by decrease of MSDs as increasing $\kappa$.}
%\textRemoved{This is a new feature that only the tracer of $\sigma_\text{tr}=6$ reveals. The behavior can be understood with the profile of potential of mean force, shown in Fig.~\ref{label:FigureA02}b. We observe that the width of $U(x)$ changes with $\kappa$. Notably, the potential width is narrower with increasing the bending stiffness until $\kappa\approx 20~k_B T$ while it becomes broader with $\kappa$ beyond as ch is maximized at $\kappa=20~k_B T$, in consistency with the magnitude of short-time MSD($\kappa$).}

Similarly to the tracer of  $\sigma_{\mathrm{tr}}=5$,
the observed bounce-back dynamics from the oscillatory behavior of the short-time MSD is pronounced in stiffer networks. As aforementioned, the observed oscillatory behavior is the consequence of the underdamped AOUP dynamics in a confining mesh. The bounce-back effect is found to be absent in the overdamped dynamics simulation of AOUPs in the same polymer network (Fig.~\ref{label:FigureS08}). 

When AOUPs have a sufficiently high self-propulsive force ($\mathrm{Pe}=72$), as shown in Fig.~\ref{label:Figure04}b, the hopping process starts to occur in $10 \lesssim t \lesssim 100$. At this activeness regime, three distinctive dynamics are observed: The tracer dynamics is ballistic in $t\lesssim \tau_0$ and Fickian in $t\gtrsim \langle \tau \rangle$. In the intermediate time ($\tau_0 \lesssim t \lesssim \langle \tau \rangle$), the tracer is subject to the viscoelastic feedback by the confining mesh. This results in intriguing MSD profiles, which illustrate intricate $\kappa$-dependent dynamics arising from the competition between the mechanical trapping force from the network and the self-propulsive force from the tracer. The inset (Fig.~\ref{label:Figure04}b) shows the short-time MSD. The amplitude of MSDs tends to decrease with $\kappa$.  

Figure~\ref{label:Figure04}c shows MSDs of the tracers at an extremely large Pe value ($\mathrm{Pe}=108$) in which the hopping processes become dominant. Similarly found in Fig.~\ref{label:Figure03}c, MSDs show super-diffusive-to-Fickian crossover dynamics at around the directional persistent time  $\tau_\mathrm{A}=10$. In the intermediate time, the amplitude of MSDs decreases and the exponent $\alpha(t)$ is lowered to unity as $\kappa$ increases. 

Below the MSD plots, in Figs.~\ref{label:Figure04}d--f, we show NGPs in the same time axes. (See also the van-Hove self-correlation functions in Fig.~\ref{label:FigureS09}). Figure~\ref{label:Figure04}d is for the passive tracer where no hopping events occur. We find the NGPs to oscillate in the short time, which reflects the bounce-back dynamics in the MSD. Overall, the magnitude of NGPs is small ($\sim 10^{-2}$), suggesting that the potential of mean force is expected be a harmonic function (Fig.~\ref{label:FigureA02}b).
Figure~\ref{label:Figure04}e shows the NGPs for AOUPs with $\mathrm{Pe}=72$. 
We find that the NGP has a broad peak, particularly with a shoulder for large $\kappa$. Similarly found in Fig.~\ref{label:Figure03}e, the peak time and value of the NGP increase with $\kappa$. 
For $\mathrm{Pe}=108$ (Fig.~\ref{label:Figure04}f), the peak time of the NGP becomes less $\kappa$-dependent. However, its magnitude increases with increasing $\kappa$ increases due to the responsive network. Such a tendency of NGP is similarly found in Fig.~\ref{label:Figure03}f, but the magnitude of NGP is larger by a factor of $10$ when $\sigma_\mathrm{tr}=6$, signyfing a huge difference in the tracer's dynamics between $\sigma_\text{tr}=5$ and $\sigma_\text{tr}=6$.

Figure~\ref{label:Figure04}g depicts the mean trapped time of AOUPs as a function of $\kappa$ for different $\mathrm{Pe}$.
Similarly to the case of $\sigma_\text{tr}=5$, the mean trapped time becomes shorter by larger activeness. 
Additionally, the mean trapped time increases exponentially with $\kappa$. From this relation, it can be deduced that $\ln \langle \tau \rangle \sim \kappa /(k_B T)$ in which $\kappa$ predominantly modulates an energy barrier for hopping into the nearest mesh. We also confirm that the trapped time is exponentially distributed (Fig.~\ref{label:FigureS10}), as the tracer of $\sigma_\text{tr}=5$ shows. In the following section, we shall explain the observed exponential law for the mean trapped time using a first-passage time theory of AOUPs. 

Figure~\ref{label:Figure04}h illustrates the mean-square flight length $\langle l^2 \rangle$ as a function of $\kappa$ for $\mathrm{Pe} \ge 36$. As observed in the results of $\sigma_{\mathrm{tr}}=5$, $\langle l^2 \rangle$ is nearly independent of $\kappa$ for small $\mathrm{Pe}$ values. 
On the other hand, for large $\mathrm{Pe}$ values of $>108$, multiple-mesh-hopping events freely occur (Fig.~\ref{label:FigureS11}) in which the hopping distance is on average shorter as the polymeric friction increases with the bending stiffness. 

Finally, in Fig.~\ref{label:Figure04}i, we show the long-time diffusivity as a function of $\kappa$ for different $\mathrm{Pe}$. 
The simulation results (symbols) and the theoretical prediction [Eq.~\eqref{eq:DL}] depicted by the solid lines are compared, showing a good agreement (The mean flight times $\langle \tau_\text{fl} \rangle$ in Fig.~\ref{label:FigureA03}b).
When $\mathrm{Pe}$ is small ($\mathrm{Pe}=54$), $\langle \tau \rangle$ dominates over $\langle \tau_\text{fl} \rangle$ and $\langle l^2 \rangle$ is independent of $\kappa$, resulting in an exponentially decaying $D_L$. For large $\mathrm{Pe}$, the exponential tendency is disrupted because $\kappa$-dependent $\langle l^2 \rangle$ and $\langle \tau_\text{fl} \rangle$ come into play.

\begin{figure*}
\includegraphics[width=0.7\textwidth]{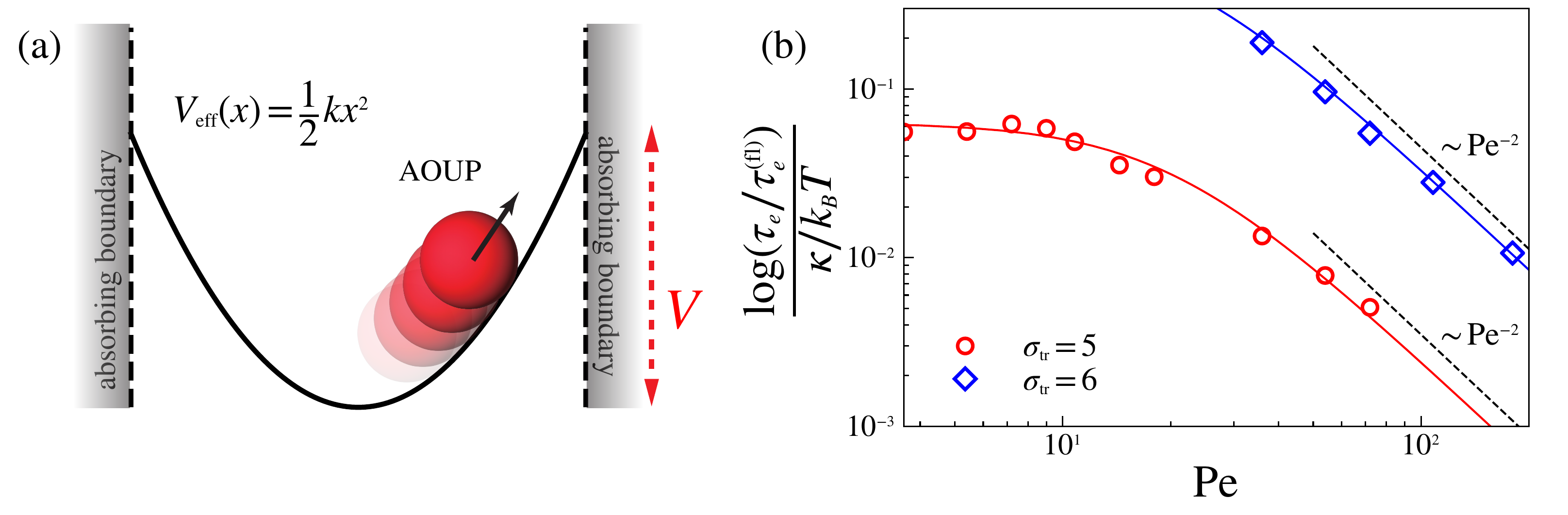}
\caption{\label{label:Figure05} 
(a) Illustration of AOUPs in a harmonic potential. When AOUPs cross the barrier of height $V$, it is considered hopping into the nearest mesh. 
(b) The plot of the rescaled mean trapped time, $\frac{\log\qty(\tau_e \slash \tau_e ^{(\mathrm{fl})})}{\kappa \slash k_B T}$, as a function of $\mathrm{Pe}$. The symbols are from our simulation data and the solid lines are from Eq.~\eqref{Eqn::C_Pe}. For $\sigma_{\mathrm{tr}}=5$ (red line), the fitting parameters are $V_1 = 0.063~k_B T$ and $ k = 15.55~k_B T\slash \sigma_0 ^2 $. For $\sigma_{\mathrm{tr}}=6$ (blue line), the fitting parameters are $V_1 = 0.84~k_B T$ and $k = 11.19~k_B T \slash \sigma_0 ^2$.
} 
\end{figure*}

\subsection{Active escape dynamics}\label{sec3D}
The above trapped-and-hopping diffusion can be conceptualized as an escaping process of AOUPs from a confining potential and hopping to the nearest mesh. 
In this context, we explain the mean trapped times $\langle \tau \rangle$ in terms of a barrier-crossing time ($\tau_e$) of a diffusing particle within a confining potential.
As supported by the potential of mean force (Fig.~\ref{label:FigureA02}), we approximate the confining potential as a harmonic potential within a minimal model framework.
We then calculate the mean-first passage time of an AOUP under two absorbing boundaries located at the mesh boundary. 
From this perspective, we present the analysis of active escape dynamics from the analogy of harmonic potential, as depicted in Fig.~\ref{label:Figure05}a.

The overadmped AOUP dynamics in a 1-D harmonic potential $V_{\mathrm{eff} }(x)=\tfrac{1}{2} k x^2$ is described by the following Langevin equation
\begin{equation}
\begin{aligned}
\gamma \frac{dx(t)}{dt} & = -k x(t) + \xi_{\mathrm{th}}(t) + F_{\mathrm{A}}(t) 
%        \\ \frac{dF_{\mathrm{A}}(t)}{dt} & = -\frac{1}{\tau_\mathrm{A}} F_{\mathrm{A}} (t) + \sqrt{\frac{\gamma^2 v_p ^2}{\tau_\mathrm{A}}} \xi_{\mathrm{A}}(t)
    \end{aligned}
\end{equation}
and, equivalently, by the two-variable ($x$ and $F_\mathrm{A}$) Fokker--Planck equation
%AOUP's dynamics is By solving two-variable ($x$ and $F_\mathrm{A}$) Fokker--Planck equation,
\begin{equation}
    \frac{\partial p}{\partial t} = 
  \frac{1}{\gamma} \pdv{}{x} [k x 
  - F_{\mathrm{A}} ]p +  \frac{k_B T}{\gamma}  \pdv[2]{p}{x}  + \frac{1}{\tau_{\mathrm{A}}} \pdv{}{F_{\mathrm{A}}} \qty[ F_\mathrm{A} p] + \frac{\gamma ^2 v_p ^2 }{\tau_\mathrm{A}} \pdv[2]{p}{F_\mathrm{A}}. 
\end{equation}
Here, $p(x, F_\mathrm{A}, t)$ is the probability density that the position and the active force of the AOUP are found to be $x$ and $F_\mathrm{A}$ at time $t$, respectively. 
Because $\xi_{\mathrm{th}}$ and $F_{\mathrm{A}}$ are both Gaussian noises, the stationary distribution for the above Fokker-Planck equation is given by a bivariate Gaussian distribution, i.e., $p_\mathrm{st}(x, F_{\mathrm{A}}) \sim \exp \qty[ - \tfrac{1}{2}
(x, F_{\mathrm{A}}) \mathbf{\Sigma}^{-1} (x, F_{\mathrm{A}})^T
]$ where $\mathbf{\Sigma}$ is the covariance matrix.
From this, we obtain the expression for the steady-state distribution of the position 
\begin{equation}\label{eq:pst}
    p_\mathrm{st}(x) \sim \exp \qty( - \frac{ \frac{1}{2} k x^2 }{k_B T +  \frac{\gamma^2 \tau_\mathrm{A} v_p^2 }{k \tau_\mathrm{A} + \gamma} }  ) = \exp\qty( - \frac{ \tfrac{1}{2} kx^2}{k_B T_{\mathrm{eff}}}).
\end{equation}
The last expression allows us to introduce the effective temperature of an AOUP 
\begin{equation}
k_B T_{\mathrm{eff}}  
= k_B T + \frac{ (k_B T)^2 \tau_\mathrm{A} }{
\sigma_{\mathrm{tr}}^2 (k \tau_\mathrm{A} + \gamma )} \mathrm{Pe}^2,
\end{equation}
consistent with the expression in Ref.~\cite{szamel2023single}.
Equation~\eqref{eq:pst} suggests that the AOUP's asymptotic long-time confinement dynamics can be mapped onto that of a Brownian particle in a harmonic potential at $T=T_{\mathrm{eff}}$. Our analytic calculation shows that the barrier-crossing time $\tau_e$ of an AOUP in the harmonic potential $V_{\mathrm{eff}}(x)$ is the Kramers time with the effective temperature:
\begin{equation}\label{eq:taue}
    \tau_e \sim \exp \left( \frac{V}{k_B T_{\mathrm{eff}}} \right)= \exp \left(
    \frac{V}{
    k_B T \qty(1 +   \frac{ k_B T \tau_\mathrm{A} }{
\sigma_{\mathrm{tr}}^2 (k \tau_\mathrm{A} + \gamma )} \mathrm{Pe}^2
)    } 
    \right)
\end{equation}
where $V$ is the potential barrier height at the absorbing boundaries (Fig.~\ref{label:Figure05}a). The above expression advances the result in Ref.~\cite{woillez2020active} in that it explicitly contains the role of the ambient heat bath ($T$).

In our framework, the details of the effective potential $V$ are modulated by the system's physical properties, such as the bending stiffness and the mesh-to-particle size ratio. For a small bending regime, the effective potential can be reasonably approximated to 
\begin{equation}\label{eq:V}
    V \simeq V_0 + V_1 \frac{\kappa}{k_B T}
\end{equation}
where $V_0$ is the potential for a flexible polymer network and the next term is responsible for the potential increase when the polymer network gains stiffness with a bending modulus $ \kappa$. Combining Eqs.~\eqref{eq:taue} and \eqref{eq:V}, we obtain the barrier-crossing time $\tau_e$ as follows:
%\begin{equation}
%    \tau_e \sim \exp \left(
%    \frac{V_0 + V_1 \frac{\kappa}{k_B T} }{
%    k_B T \qty(1 +   \frac{ k_B T \tau_\mathrm{A} }{
%\sigma_{\mathrm{tr}}^2 (k \tau_\mathrm{A} + \gamma )} \mathrm{Pe}^2
%)} \right).
%\end{equation}
\begin{equation}\label{tau_e}
\tau_e = \tau_e^{\mathrm{(fl)}}\exp \left( \frac{V_1}{k_B T \qty( 1+\frac{\tau_\mathrm{A}k_BT}{\sigma_{\mathrm{tr}}^2(k\tau_\mathrm{A}+\gamma)}\mathrm{Pe}^2 )}\frac{\kappa}{k_BT}\right)
\end{equation}
where $\tau_e^{\mathrm{(fl)}}$ is the mean escape time in the limit of a flexible polymer network ($\kappa\to0$). Note that Eq.~\eqref{tau_e} includes two undetermined potential constants of a given polymer network: $V_1$ and the spring constant $k$, which depend on some physical parameters, e.g., the mesh-to-particle size ratio. Because both constants are independent of $\kappa$, we anticipate that the mean escape (i.e., trapped) time of an AOUP in a polymer network follows an exponential law of the form:
\begin{equation}\label{tau_e2}
\tau_e = \tau_e^{\mathrm{(fl)}}\exp\left(C\frac{\kappa}{k_BT}\right)
\end{equation}
where the prefactor $C = C(\mathrm{Pe},\sigma_\mathrm{tr})$ reads
\begin{equation}
    C(\mathrm{Pe}, \sigma_{\mathrm{tr}}) 
        = \frac{\log (\tau_e/\tau_e ^{\mathrm{(fl)}}) }{\kappa \slash k_B T }
    = 
    \frac{V_1}{
k_B T\qty(    1 +   \frac{ k_B T \tau_\mathrm{A} }{
\sigma_{\mathrm{tr}}^2 (k \tau_\mathrm{A} + \gamma )} \mathrm{Pe}^2
    )}
\label{Eqn::C_Pe}
\end{equation}
explaining the stiffness effects on the active escaping dynamics.
Our theory, Eq.~\eqref{tau_e2}, describes the exponential growth law of the mean trapped times of AOUPs observed in our simulation (Figs.~\ref{label:Figure03}g and \ref{label:Figure04}g).  

We obtain two important properties of the active escaping process from Eq.~\eqref{tau_e}. 
First, the effect of bending stiffness, $C(\mathrm{Pe}, \sigma_{\mathrm{tr}})$,
illustrates different behaviors on small and large $\mathrm{Pe}$ regimes. When $\mathrm{Pe}$ is small, the active escaping dynamics barely depends on the network stiffness (i.e., $C \simeq \frac{V_1}{k_B T}$). However, when $\mathrm{Pe}$ is increased, the stiffness factor $C$ decays as $C \sim {\mathrm{Pe}^{-2}}$ with $\mathrm{Pe}$. 
In Fig.~\ref{label:Figure05}b, we plot the mean trapped time in the form of $\frac{\log\qty(\tau_e \slash \tau_e ^{(\mathrm{fl})})}{\kappa \slash k_B T}$ as a function of $\mathrm{Pe}$ from our simulation data, along with Eq.~\eqref{Eqn::C_Pe} [solid line]. 
When $\sigma_{\mathrm{tr}}=5$, the theoretical line is plotted with $V_1$ determined from the Brownian case ($\mathrm{Pe}=0$) in Fig.~\ref{label:Figure03}g and $k$ being a fitting parameter.
When $\sigma_{\mathrm{tr}}=6$, since a hopping event is rarely observed during our simulation time, we fit the data with Eq.~\eqref{Eqn::C_Pe}, with $V_1$ and $k$ being the fitting parameters. Our simulation results in Fig.~\ref{label:Figure05}b indeed demonstrate that the stiffness factor $C$ illustrates the theoretically expected two scaling behaviors. 

%As explained in Eq.~\eqref{eq:V}, 
The parameter $V_1(\sigma_{\mathrm{tr}})$ quantifies the slope of potential increase as the polymer network stiffens. 
Our above analysis reveals that $V_1 (\sigma_{\mathrm{tr}} = 5) = 0.063~k_B T$ and $V_1(\sigma_{\mathrm{tr}}=6)=0.84~k_B T$. 
Note that $V_1(\sigma_{\mathrm{tr}}=6) \gg V_1 (\sigma_{\mathrm{tr}} = 5)$, reflecting that the effect of polymer stiffness on the active escape dynamics is dramatically amplified as a tracer is larger. Moreover, as Fig.~\ref{label:Figure05}b shows, the stiffness effect is larger than an order of magnitude for the tracer size of $\sigma_{\mathrm{tr}}=6$ than $\sigma_{\mathrm{tr}}=5$.
This also confirms that the stiffness effect on the active escape is dramatically increased when the tracer size increases.

%From the data in Fig.~\ref{label:Figure03}g and Fig.~\ref{label:Figure04}g, we obtain $C(\mathrm{Pe})$ by fitting the trapped times, $\langle \tau \rangle$.
%So, $C = \frac{d}{d\kappa} \log \langle \tau \rangle$.
%It is represented in Fig.~\ref{label:Figure05} we fit this by Eq.~\eqref{Eqn::C_Pe}. (For $\sigma_{\mathrm{tr}}=5$, $V_1$ is from the data for passive tracers and $k$ is the fitting parameter. For $\sigma_{\mathrm{tr}}=6$, both $V_1$ and $k$ are fitting parameters.)

%In this study, the stiffness of the polymer network, $\kappa$ modulates the energy barrier, and the mean trapped times $\langle \tau \rangle$ follows the exponential law as in Fig.~\ref{label:Figure03}~and~\ref{label:Figure04}. 
%Also, the fitting parameter is in Fig.~\ref{label:Figure05}. When the activeness is small, as in inset of Fig.~\ref{label:Figure05}, $C(\mathrm{Pe})$ remains constant of value $\approx 0.06$ and when $\mathrm{Pe}$ is large, the stiffness effect can be described by the power-law $C(\mathrm{Pe}) \sim \mathrm{Pe} ^{-2}$.

%Note that $C(\mathrm{Pe})$ describes the stiffness effects on the polymer network. 
%It is larger than an order of magnitude for the tracer size $\sigma_{\mathrm{tr}}=6$ than $\sigma_{\mathrm{tr}}=5$.
%This implies that the stiffness effect on active escape is dramatically increased when the tracer size increases.

\section{discussion}\label{sec4}
\begin{figure*}    \includegraphics[width=0.95\textwidth]{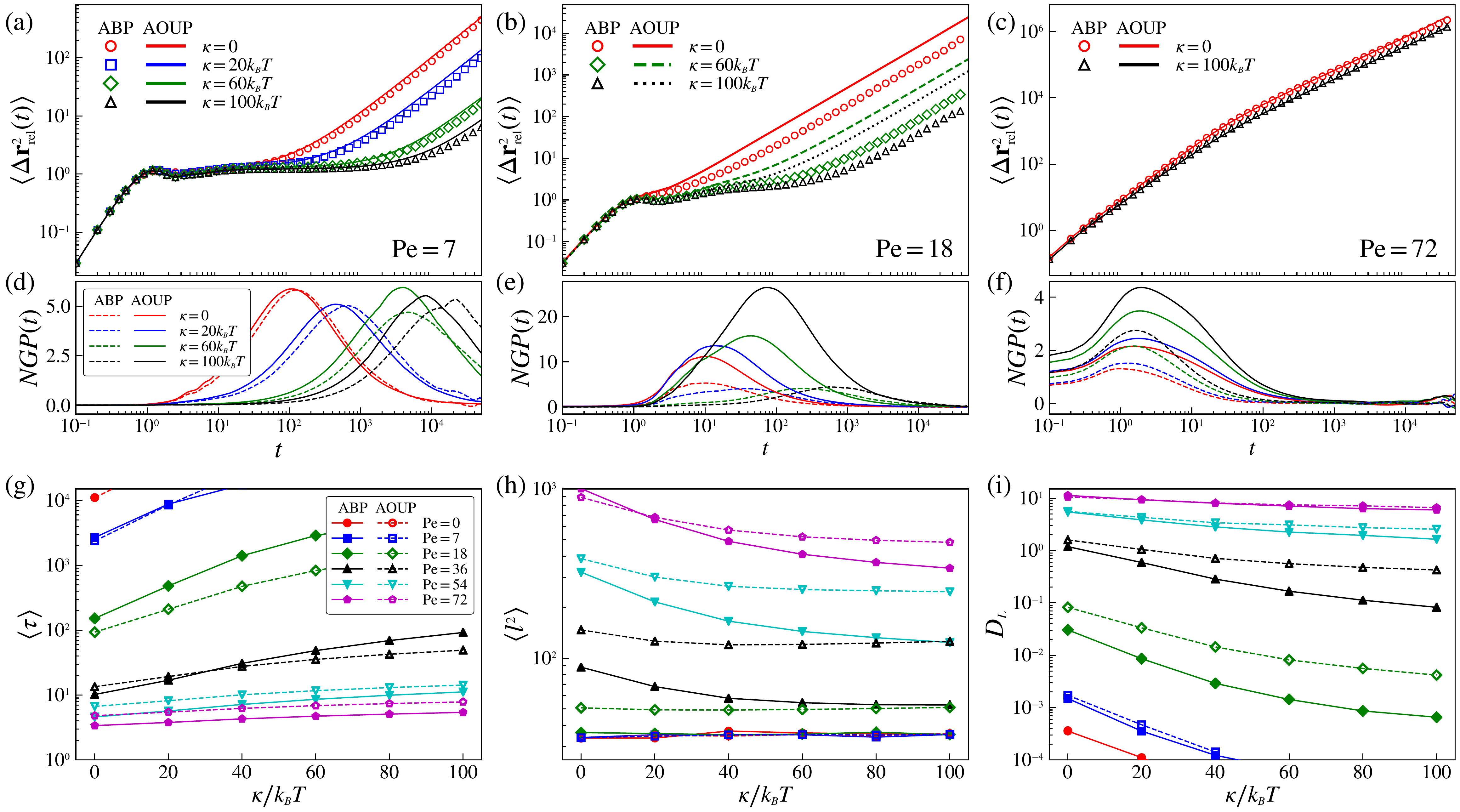}
\caption{\label{label:Figure06}
The simulation results for ABP tracers ($\sigma_{\mathrm{tr}}=5$) in semi-flexible polymer networks in comparison with those of AOUPs of the same size (Fig.~\ref{label:Figure03}). MSD for (a) $\mathrm{Pe}=7$, (b) $\mathrm{Pe}=18$, and (c) $\mathrm{Pe}=72$.
NGP for (d) $\mathrm{Pe}=7$, (e) $\mathrm{Pe}=18$, and (f) $\mathrm{Pe}=72$.
(g) Mean trapped time $\langle \tau \rangle$, (h) mean-square flight length $\langle l^2 \rangle$, and (i) long-time diffusivity $D_L$ as a function of $\kappa$, respectively. 
}
\end{figure*}

We studied the diffusion dynamics of active tracer particles in a semi-flexible polymer network by performing underdamped Langevin dynamics simulations. The observed tracer dynamics spans a wide time scale, ranging from the short-time ballistic dynamics to the long-time diffusive dynamics. 
For mesh-sized tracers that are tightly trapped in the network, we observed the bounce-back dynamics in the short-time regime~\cite{nguyen2021active, caprini2021inertial, lowen2020inertial}, manifested by oscillations in MSDs and NGPs (Figs.~\ref{label:Figure03}~and~\ref{label:Figure04}).
This unique behavior was further validated in Figs.~\ref{label:FigureS04}~and~\ref{label:FigureS08}, where the overdamped Langevin dynamics simulations of the same system produced the MSD without the bounce-back oscillations.

In our study, we adopted the AOUP model for the active tracer. 
As described in Introduction, there exist other relevant active particle models, including the active Brownian particle (ABP) model~\cite{howse2007self} or the run-and-tumble particle model~\cite{berg2004coli}. 
In these models, the active force is consistently characterized by an exponentially decaying directional memory~\cite{mori2021condensation, sarkar2023stationary}, but their microscopic mechanism is distinct: 
The ABP moves with a constant propulsion speed while its orientation randomly changes.
The run-and-tumble particle follows straight paths with intermittent tumbling events, reflecting a distinct succession of straight-line segments. It remains an open question to systematically examine how the active diffusion in a polymer network depends on these active tracer models.

To obtain an insight into the dependence of active tracer models on the active diffusion reported in Sec.~\ref{sec3}, we repeated the underdamped Langevin dynamics simulations of our semi-flexible polymer networks with the ABP model. 
For a direct comparison with our AOUP model, we neglected the rotational inertia and set the rotational diffusivity to $D_R = 1 /(2 \tau_\mathrm{A})$, which yields the same propulsion memory and MSDs of the AOUP described in Eqs.~\eqref{eqn_noisecorr} and \eqref{msdfree}, respectively.

%The detailed comparison is presented in Fig.~\ref{label:Figure06}. 
In Fig.~\ref{label:Figure06}a, we compare the MSDs of ABPs and AOUPs at $\mathrm{Pe}=7$. When $\mathrm{Pe}$ is relatively small compared to the potential barrier to escape from the trapping, the effect of self-propulsive movement is negligible and the MSDs for both ABP and AOUP tracers are almost identical. 
However, as the activeness is increased to $\mathrm{Pe}=18$ (Fig.~\ref{label:Figure06}b), we can observe the difference between the two models. Here, the microscopic features of each model play a distinct role in the MSDs. 
While the MSDs of the two models coincide in the short-time regime where the tracer rarely interacts with the polymer network, 
the AOUPs in the intermediate-to-long-time regimes consistently display larger MSDs compared to ABPs. This indicates that AOUPs have a greater propensity to escape from trapping than ABPs. 
Notably, when the activeness is further increased to $\mathrm{Pe}=72$ (Fig.~\ref{label:Figure06}c), the MSDs of both models converge, reflecting a scenario where trapping occurrences are rare, and ABPs and AOUPs explore the network with a similar self-propulsion tendency.
In Figs.~\ref{label:Figure06}d--f, we analyze the NGPs. Consistent with the MSDs, non-Gaussianity of the active movement is alike for both models at $\mathrm{Pe}=7$ while the AOUP exhibits stronger non-Gaussian diffusion than the ABP at high $\mathrm{Pe}$ numbers. 
In Fig.~\ref{label:Figure06}g, we investigate the mean trapped times ($\langle \tau \rangle$) for both models. The $\langle \tau \rangle$ is similar to each other at $\mathrm{Pe}=7$. For intermediate $\mathrm{Pe}$ values, the AOUP has a shorter mean trapped time than the ABP. When $\mathrm{Pe}$ is increased further, the trapping effect is more negligible, resulting in a smaller difference in $\langle \tau \rangle$ between the two models.  For all $\mathrm{Pe}$ values explored, $\langle \tau \rangle$ of the ABP is shown to increase with $\kappa$, mirroring the behavior observed in the AOUP.
Figure~\ref{label:Figure06}h shows the mean square flight lengths of ABPs and AOUPs. The $\langle l^2 \rangle$ of AOUPs is larger than that of ABPs at intermediate $\mathrm{Pe}$ values, with the disparity becoming less dominant for larger $\mathrm{Pe}$.
Finally, Fig.~\ref{label:Figure06}i presents the long-time diffusivity of ABPs and AOUPs. Consistent with other observables, the difference between ABPs and AOUPs is noticeable in the regime of intermediate $\mathrm{Pe}$ and becomes negligible at small and large $\mathrm{Pe}$ values.
%In summary, ABPs exhibit qualitatively similar trapped-and-hopping active diffusion as AOUPs, as observed in our main study. The distinction in their active diffusion becomes evident primarily in the intermediate $\mathrm{Pe}$ regime, where the active escaping dynamics from a trapped site are influenced by the microscopic self-propulsion mechanisms inherent to each model.
In a nutschell, ABPs exhibit qualitatively similar trapped-and-hopping active diffusion as AOUPs, as observed in our main study. The distinction in their active diffusion is evident primarily in the intermediate $\mathrm{Pe}$ regime, where the active escaping dynamics from a trapped site are largely dominated by the microscopic self-propulsion mechanism inherent to each model. 

%[WHY?]A recent study reported the dynamics of ABPs in a polymer network that is based on a diamond lattice~\cite{kumar2023dynamics}, in which the high activeness and the large tracer size yield the active diffusion independent of the polymer stiffness. The studies show that the structure of the polymer network is also an important factor determining the active tracers' dynamics. 

%\cmtWK{Rewrite the summary of this paper here.}
%They investigate the effects of self-propulsion force, particle size, sticky interactions, and stiffness on MSD, its exponent, and van-Hove distribution. As in our study, their results show the suppression of tracer dynamics by the stiffness. However, their finding reveals that the stiffness effects wash out when the activity is high and the tracer size is larger than the mesh size, while our study reveals that increasing stiffness decreases superdiffusive exponents and long-time diffusivity. This difference may originate from the structures of the polymer network.

\section{conculsions}\label{sec5}
We have simulated AOUPs in a semi-flexible polymer network and systematically investigated the effects of bending stiffness on the trapped-and-hopping active diffusion.
We have found that the diffusivity of AOUPs is mainly governed by the particle-to-mesh size ratio and semi-flexibility of the polymer network as well as $\mathrm{Pe}$.

When the tracer size is marginally smaller than the mesh size, the polymer's stiffness does not drastically affect the trapped-and-hopping dynamics. This is because the trapping force is negligibly small while hopping frequently occurs in the presence of minor obstacle effects from the network.   
However, when the tracer size is comparable to the mesh size, the stiff polymer network hinders the hopping and enhances the trapped time.
Consequently, the mean trapped time becomes longer as the polymer becomes stiffer. However, the flight length is a decreasing function of the bending stiffness only for large $\mathrm{Pe}$, in which the long-jump events are more hindered by the stiff polymers. 
For active tracers marginally larger than the mesh size, we have observed that the tracers suffer confined motions easily when the activeness is weak. When the activeness increases, the hopping process starts to occur, by which the trapped time depends on the polymer stiffness sensitively while the flight length does not. However, for largely activated tracers, the hopping dynamics becomes dominant and the flight length changes sensitively with varying bending stiffness, while the trapped time becomes less sensitive.

%\textAdded{[What is the important point in the first-passage part?] We showed that our simulation results has a good agreement with theory of active escape dynamcis in the harmonic potential. The interplay of polymer network stiffness and tracer's activeness and size is successfully interpreted via the analogy of active escape dynamics in harmonic potential. Our results We reported that the effect of stiffness on active escaping dynamics could differ at small and large $\mathrm{Pe}$ and the stiffness effects are dramatically amplified with an increase in tracer size.} 
%\textAdded{We also developed the theory of active escape dynamics in harmonic potential. Our simulation results which shows the interplay of polymer network stiffness and tracer's activeness and size, has a good agreement with the developed theory of AOUPs in harmonic potential. }

Our study, which was extended from the previous work for flexible polymer networks~\cite{kim2022active}, demonstrates that the polymer's semi-flexibility plays an important role in sensitively modulating the active diffusion of the mesh-sized tracers confined in the polymer network. The observed trapped-and-hopping mechanism thus provides insights to better understand the active dynamics under confinement. 

%\iffalse
\begin{acknowledgments}
%\begin{acknowledgement}
%This work was supported by the National Research Foundation (NRF) of Korea (No.~2020R1A2C4002490).
This work was supported by the National Research Foundation of S. Korea via No.~RS-2023-00218927.
W.K.K. acknowledges the financial support from the KIAS Individual Grants (CG076002) at Korea Institute for Advanced Study.
We acknowledge the Center for Advanced Computation at Korea Institute for Advanced Study for providing computing resources for this work.
\end{acknowledgments}
%\end{acknowledgement}
%\fi

%\begin{suppinfo}
%\end{suppinfo}

% %=====================================FIG 3=========================================================
% %===================================================================================================
% \begin{figure*}
%     \includegraphics[width=12cm]{supp1.pdf}
%     \caption{(a) The mean square displacement and (b--d) non-Gaussian parameters for small passive (b: $\mathrm{Pe}=0$) and active (c: $\mathrm{Pe}=1$ and d: $\mathrm{Pe}=5$) tracers ($\sigma_{\mathrm{tr}}=1$). 
%     \label{small_msdngp}}
% \end{figure*}
% %===================================================================================================
% %===================================================================================================

\section*{APPENDIX}
\appendix

\renewcommand{\thefigure}{A\arabic{figure}}
\renewcommand{\thesubsection}{\Alph{subsection}}

\setcounter{figure}{0}

% %=====================================FIG 3=========================================================
% %===================================================================================================
% \begin{figure*}
%     \includegraphics[width=18cm]{system_4.pdf}
%     \caption{\label{fig00} 
%     (a) The full structure of the semi-flexible polymer network and (b--e) zoomed-in figures with various stiffness ($\kappa$).
%     (a) $\kappa=40$ (b) $\kappa=0$ (c) $\kappa=10$ (d) $\kappa=60$ and (e) $\kappa=100$. 
%     }
% \end{figure*}
% %===================================================================================================
% %===================================================================================================

%=====================================FIG 3=========================================================
%===================================================================================================
\begin{figure*}
    \includegraphics[width=15cm]{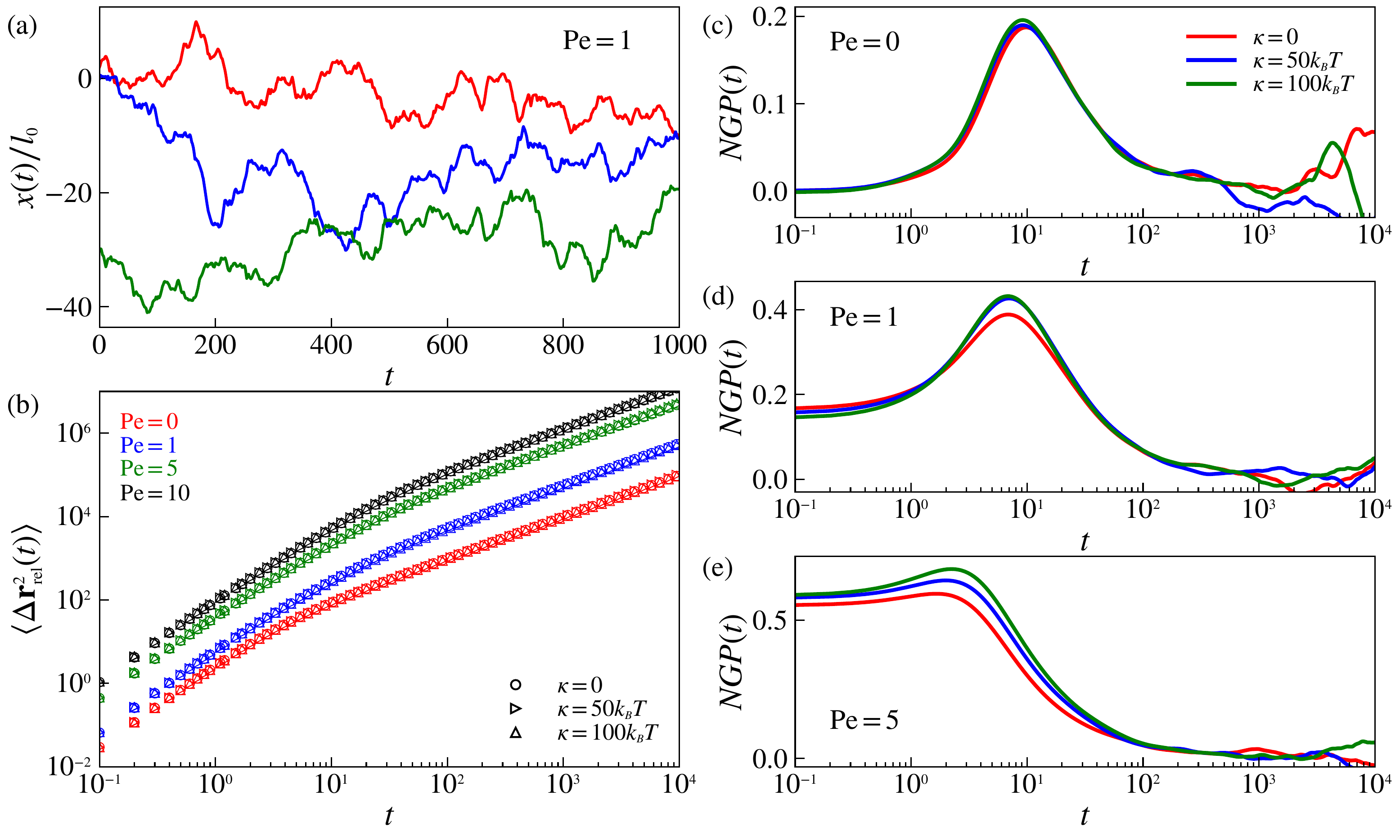}
    \caption{\label{label:FigureA01} Active diffusion for small active tracers. (a) Sample trajectory for different polymer network stiffness. 
    (b) MSD 
    (c) NGP for passive tracers
    (d,e) NGP for active ($\mathrm{Pe}=1$, $5$) tracers
    }
\end{figure*}

\begin{figure*}
\includegraphics[width=15cm]{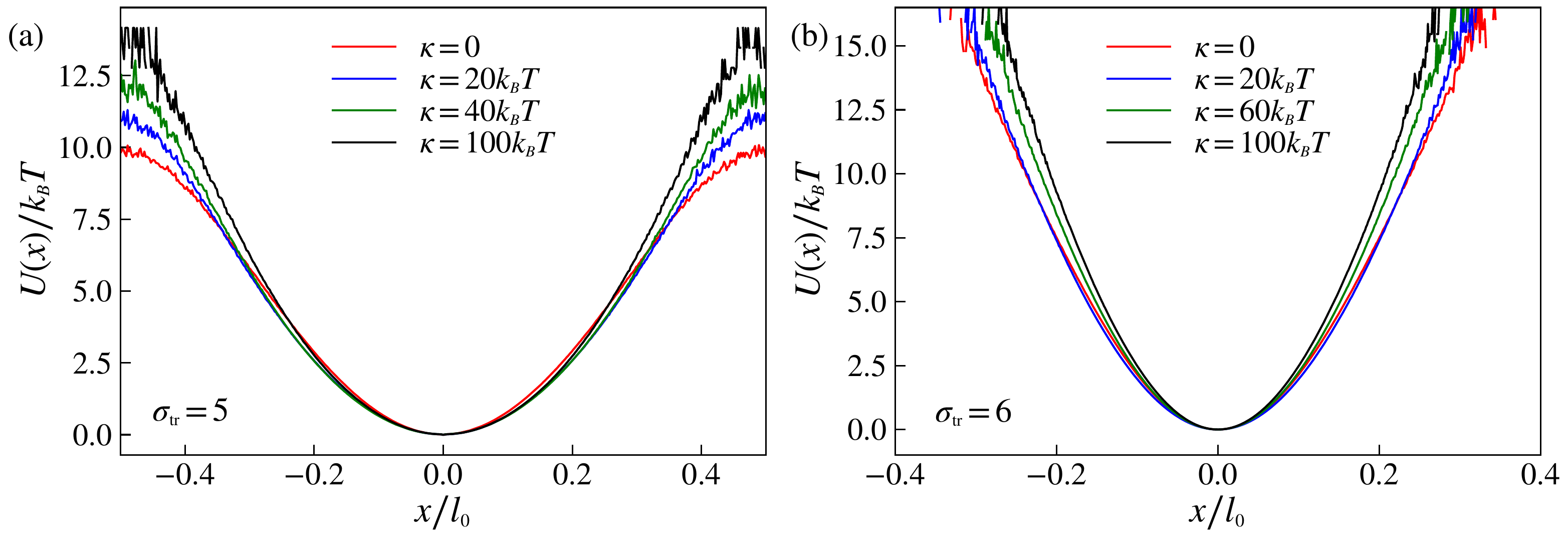}
\caption{\label{label:FigureA02}
The potential of mean forces for the tracers ($\mathrm{Pe}=0$) of size (a) $\sigma_{\mathrm{tr}} = 5$ and (b) $\sigma_{\mathrm{tr}} = 6$ in a flexible and semi-flexible polymer networks of $\kappa=0,~20 k_B T, ~60  k_B T$, and $100 k_B T$.
}
\end{figure*}

\subsection{Classification of trapped-and-hopping events}
To classify the tracer trapping and hopping in the trajectories, we first skeletonize the trajectories. 
From every relative particle positions $\mathbf{r}_{\mathrm{rel},i} (t) $, we compute the skeletonized trajectories $\hat{x} \equiv \mathcal{F}_{\mathrm{Bi}}\qty[ \mathcal{F}_{\mathrm{Bi}} \qty[x] ]$, where $\mathcal{F}_{\mathrm{Bi}}$ is the bilateral filter~\cite{tomasi1998bilateral} and $x$ is a component of the relative trajectory. This is defined as
\begin{equation}
    \mathcal{F}_{\mathrm{Bi}}[x] \equiv
    \frac{1}{W_{\mathrm{p}}} \sum_{t_i \in \Omega} x_j\left(t_i\right) f\left(\left|x_j\left(t_i\right)-x_j\left(t_k\right)\right|\right) g\left(\left|t_i-t_k\right|\right),
\end{equation}
where $W_{p}=\sum _{x_{i}\in \Omega }{f\left(\left|x_j\left(t_i\right)-x_j\left(t_k\right)\right|\right) g\left(\left|t_i-t_k\right|\right)}$ is the normalization factor, $f$ and $g$ are the Gaussian kernels given by $f(x)=e^{-x^2/\sigma_x ^2}$ and $g(t)=e^{-t^2/\sigma_t ^2}$. We set $\sigma_x = 5$ and $\sigma_t =1$. 
The bilateral filter's edge-preserving properties produce skeletonized trajectories $\hat{x}$ with noise-free trapped parts and instantaneous jumps.

Next, we identify hopping events from the filtered trajectory $\hat{x}$ by examining our criteria $\left| \hat{x}(t + 2t_0) - \hat{x}(t - t_0 )\right| > l_c$, where the cutoff is $l_c = 0.6 \times l_0 = 3$. We then determine the initial and final times for the hopping by checking if the original trajectory $x(t)$ and filtered trajectory $\hat{x}$ share the same direction in one component. Specifically, if four segments $x(t+t_0) - x(t)$, $x(t)-x(t-t_0)$, $\hat{x}(t+t_0)-\hat{x}(t)$, and $\hat{x}(t)-\hat{x}(t-t_0)$ have the same direction, and the tracer is in a hopping state at $t+t_0$ or $t-t_0$, we set the tracer to be in a hopping state at $t$. We iteratively apply this condition to establish the initial and final hopping times.

Since the trajectory is in 3D, we define a trapped state when the trajectory is not in a hopping state for any components. The trapped time is then defined as the duration of consecutive trapped states. To determine the flight time, we find the duration of consecutive hopping states. Finally, for the flight length, we calculate the chord length of the hopping state in 3D by using two ends of positions of the consecutive hopping interval.

\subsection{Active diffusion of small tracers}

Figure~\ref{label:FigureA01} represents the active diffusion of small tracers {($\sigma_{\mathrm{tr}} = 1$)} in the polymer networks. 
For the small tracers, the polymer network acts as a trivial obstacle to the tracers. Figure~\ref{label:FigureA01}a is the sample trajectory of small active tracers ($\mathrm{Pe}=1$). 
Figure~\ref{label:FigureA01}b represents the MSDs for small active tracers, where the color represents $\mathrm{Pe}$ and the symbols represent the different stiffness $\kappa$. The simulation results show that varying the polymer's bending stiffness does not significantly affect the diffusion dynamics of small active tracers.
Figures~\ref{label:FigureA01}c--e represent the NGPs of small active tracers. The NGPs of the small active tracers in a semi-flexible polymer network exhibit only a slight increase compared to those in a flexible polymer network.

\subsection{The potential of mean forces for tracers of size $\sigma_{\mathrm{tr}}=5$ and $6$}

For the Brownian tracer ($\mathrm{Pe}=0$) confined in a polymer mesh, we calculate the potential of mean force by the following relations:
%(Equilibrium distributions of passive particles in potential).
\begin{equation}
\begin{aligned}
    p(x) & \sim e^{-\beta U(x)}  \\
    U(x) & = - k_B T \ln (p(x)) + k_B T \ln \qty(p_0)
\end{aligned}
\end{equation}
Here, for convenience, we set $U(0)=0$ by adding the second term where $p_0 \equiv p(x=0)$.

Figure~\ref{label:FigureA02} depicts the potential of mean force calculated from the trajectories of the Brownian tracers, which shows how the potential of mean force is modulated by the polymer's stiffness.

\begin{figure*}
\includegraphics[width=15cm]{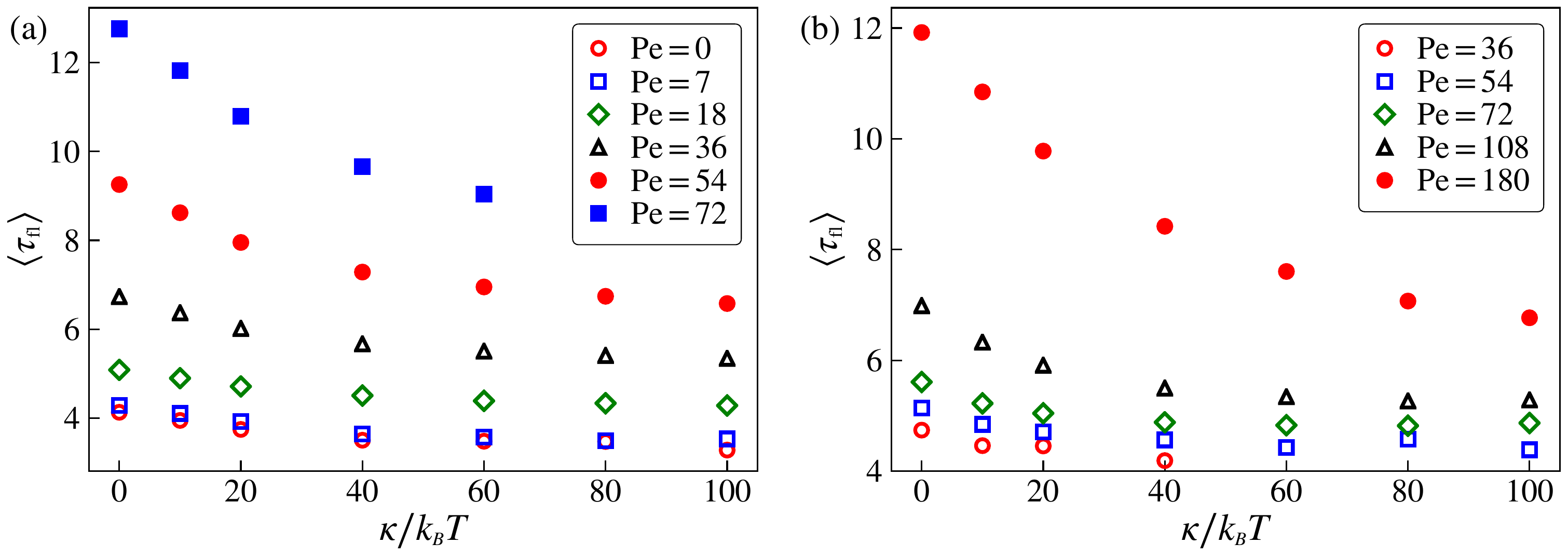}
\caption{\label{label:FigureA03}
Mean flight time ($\tau_\text{fl}$) of AOUPs of size (a) $\sigma_\text{tr}=5$ and (b) $\sigma_\text{tr}=6$  as a function of $\kappa$ for various $\mathrm{Pe}$ values.
}
\end{figure*}

\subsection{The mean flight times for tracers of size $\sigma_{\mathrm{tr}}=5$ and $6$}

Figure~\ref{label:FigureA03} shows the mean flight time as a function of $\kappa$ for the active tracers with various Pe values. 

\bibliography{ref}% Produces the bibliography via BibTeX.

\end{document}